\documentclass[aps,reprint,longbibliography]{revtex4-1}

\usepackage{lipsum}
\usepackage{graphicx}
\usepackage{subfigure}
\usepackage{color}
\usepackage{amsmath}
\usepackage[linktocpage,colorlinks=true,linkcolor=blue,citecolor=blue,urlcolor=blue,breaklinks=true]{hyperref}
\usepackage{verbatim}
\usepackage{breakcites}
\usepackage{wrapfig}
\usepackage{soul}

\begin{document}

\title{Universal entrainment mechanism controls contact times with motile cells}

\author{Arnold J. T. M. Mathijssen}
\thanks{A.M. and R.J. contributed equally to this work and are joint lead authors. \ar{A.M., R.J. and M.P. designed the study, analysed results, developed the theory, wrote the manuscript; R.J. performed the experiments; A.M. developed the model, performed simulations.}}
\affiliation{Department of Bioengineering, Stanford University, 443 Via Ortega, Stanford, CA 94305, United States}

\author{Rapha\"{e}l Jeanneret} 
\thanks{\ar{ R.J. Current Address: IMEDEA,
University of the Balearic Islands,
Carrer de Miquel Marqu\`es, 21
07190 Esporles,
Illes Balears}}
\affiliation{Physics Department, University of Warwick, Gibbet Hill Road, Coventry CV4 7AL, United Kingdom}

\author{Marco Polin}
\email[Correspondence: ]{M.Polin@warwick.ac.uk}
\affiliation{Physics Department, University of Warwick, Gibbet Hill Road, Coventry CV4 7AL, United Kingdom}

\newcommand\mummy{\nobreak\mbox{~$\mu$m} }
\newcommand\mummys{\nobreak\mbox{~$\mu$m.s$^{-1}$}}
\renewcommand{\vec}[1]{\boldsymbol{#1}}
\newcommand{\bnabla}{\vec{\nabla}}
\newcommand{\RS}{r_\textmds{S}}
\newcommand{\RT}{r_\textmds{P}}
\newcommand{\VS}{v_\textmds{S}}
\newcommand{\VT}{v_\textmds{P}}
\newcommand{\impar}{b}
\newcommand{\actson}{,}
\newcommand{\chlamy}{\textit{Chlamydomonas }}
\newcommand{\ud}{\mathrm{d}}
\newcommand{\textmds}[1]{\textmd{\tiny{#1}}}
\newcommand{\fig}[1]{Fig.~\ref{#1}}
\newcommand{\eq}[1]{Eq.~\ref{#1}}
\newcommand{\sect}[1]{\S\ref{#1}}
\newcommand{\tbl}[1]{Table~\ref{#1}}
\newcommand{\kbt}{k_\textmd{B}\textmd{T}}
\newcommand{\ar}[1]{\textcolor{black}{#1}}

\date{\today}

\begin{abstract}
\noindent
Contact between particles and motile cells underpins a wide variety of biological processes, from nutrient capture and ligand binding, to grazing, viral infection and cell-cell communication. The window of opportunity for these interactions depends on the basic mechanism determining contact time, which is currently unknown. By combining experiments on three different species -{\it Chlamydomonas reinhardtii}, {\it Tetraselmis subcordiforms}, and {\it Oxyrrhis marina}- simulations and analytical modelling, we show that the fundamental physical process regulating proximity to a swimming microorganism is hydrodynamic particle entrainment. The resulting \ar{distribution of} contact times \ar{is derived} within the framework of Taylor dispersion as a competition between advection by the cell surface and microparticle diffusion, and predict\ar{s} the existence of an optimal tracer size \ar{that} is also observed experimentally. Spatial organisation of flagella, swimming speed, swimmer and tracer size influence entrainment features and provide trade-offs that may be tuned to optimise \ar{the estimated probabilities for} microbial interactions like predation and infection.
\end{abstract}

\maketitle

\section{Introduction}

The wide variety of microbial interactions is often deeply influenced by physics. Within biofilms, electric currents can coordinate cellular metabolic rates \cite{Prindle2015}, while wrinkles draw nutrients by capillarity \cite{Wilking2013}. Microscopic flow fields \cite{Drescher2010f,Guasto2010,Drescher2011} can lead to large-scale collective motion \cite{Zhang2010,Cisneros2010,Sokolov2012} with enhanced drug resistance \cite{Lai2009,Butler2010}, surprising rheological properties \cite{Sokolov2009, Rafai2010,Lopez2015}, and global features controllable by structured confinement \cite{Wioland2013,  mathijssen2015hotspots, Ravnik2013, Lushi2014, mathijssen2015hydrodynamics}. When coupled with population-wide taxis, these flows result in macroscopic instabilities \cite{Platt1961,Plesset1974,Pedley1988} which increase nutrient fluxes \cite{Tuval2005} and can \ar{provide} unexpected new avenues for capture and manipulation of small objects \cite{Dervaux2016}.

For swimming microorganisms, many interactions hinge on close contact. \ar{These include fundamental processes like nutrient uptake \cite{Magar2003, Michelin2011, Tam2011, doostmohammadi12, lambert13, ishikawa2016nutrient, dolger2017swimming,Short2006}; viral and fungal infection of microorganisms of ecological and commercial importance \cite{Seisenberger2001, Taylor2013,carney14}, eukaryotic fertilisation \cite{riffell2007}; and grazing, which happens on natural preys \cite{roberts2010feeding, davidson2011oxyrrhis, gilpin2016vortex, dolger2017swimming}  as well as marine microplastics \cite{cole2013microplastic, wright2013physical}, and is recently being discovered as a fundamental behaviour in many strains of motile green algae until recently regarded as exclusive phototrophs \cite{maruyama13, carney14, selosse16}.
 With the exception of complex feeding currents in ciliates like {\it Vorticella} \cite{pepper10,pepper13} or {\it Paramecium} \cite{jung14,zhang15}, the window of opportunity for these microbial interactions to take place will depend on} a finite contact time $T$.  For a constant success rate per unit time $\Omega$, the probability that the interaction is successful is given by $p(T) = 1- e^{-\Omega T}$ \cite{gilpin2016vortex}. Large values of $p(T)$ will be favoured by long contact times, and will therefore depend on the physics that regulates proximity. Although the theoretical basis of contact times is still developing \cite{Bally2011,Banerjee2016,Xu2016}, it is reasonable to expect that a key role will be played by the properties of the near field, the region close to the cell body \cite{dolger2017swimming}. 
As already noted in this context by Purcell \cite{purcell1977life}, the fluid layer close to a swimming microorganism is expected to be carried along by it. 
Small objects sufficiently close to a microswimmer are therefore entrained \cite{Darwin1953} and stay in close contact with it for the time required to escape the near-field region. Entrainment converts a temporal quantity, the contact time $T$, into a readily measurable spatial quantity, the entrainment length $L$.
While recent studies have provided numerical support for particle entrainment by microorganisms \cite{Pushkin2013, mueller2017fluid, shum2017entrainment}, experimental evidence is limited to the microalga \textit{Chlamydomonas reinhardtii} \cite{Jeanneret2016}, and it is not clear whether or not this phenomenon is a general feature of microbial motility. At the same time, there are currently no theoretical predictions for the duration of these contact events, as a clear picture of the physics underlying entrainment is lacking \cite{Lin2011, Pushkin2013, Jeanneret2016, Thiffeault2015, mueller2017fluid, shum2017entrainment}.

Here we combine experimental, numerical and theoretical approaches to investigate particle entrainment by microorganisms.
\ar{Experiments with the green microalgae \textit{Chlamydomonas reinhardtii} (CR) and \textit{Tetraselmis subcordiforms} (TS) -pulled by different numbers of anterior flagella- and the dinoflagellate \textit{Oxyrrhis marina} (OM) -pushed by a posterior flagellum- demonstrate that entrainment is indeed a robust generic feature amongst swimming cells, whose existence is independent of the propulsion strategy.
Entrainment is shown to be a direct consequence of two universal traits: advection by a no-slip cell surface, as recently suggested in \cite{mueller2017fluid}, and particle diffusion.
A first-passage Taylor-dispersion argument combines these fundamental physical ingredients, allowing for analytical estimates of the mean contact time and the full entrainment distribution, while offering an intuitive understanding of the observed existence of an optimal particle size for entrainment. We conclude by discussing potential consequences on the probability of successful interactions.}

\begin{figure}
\centering
\includegraphics[width=0.96\linewidth]{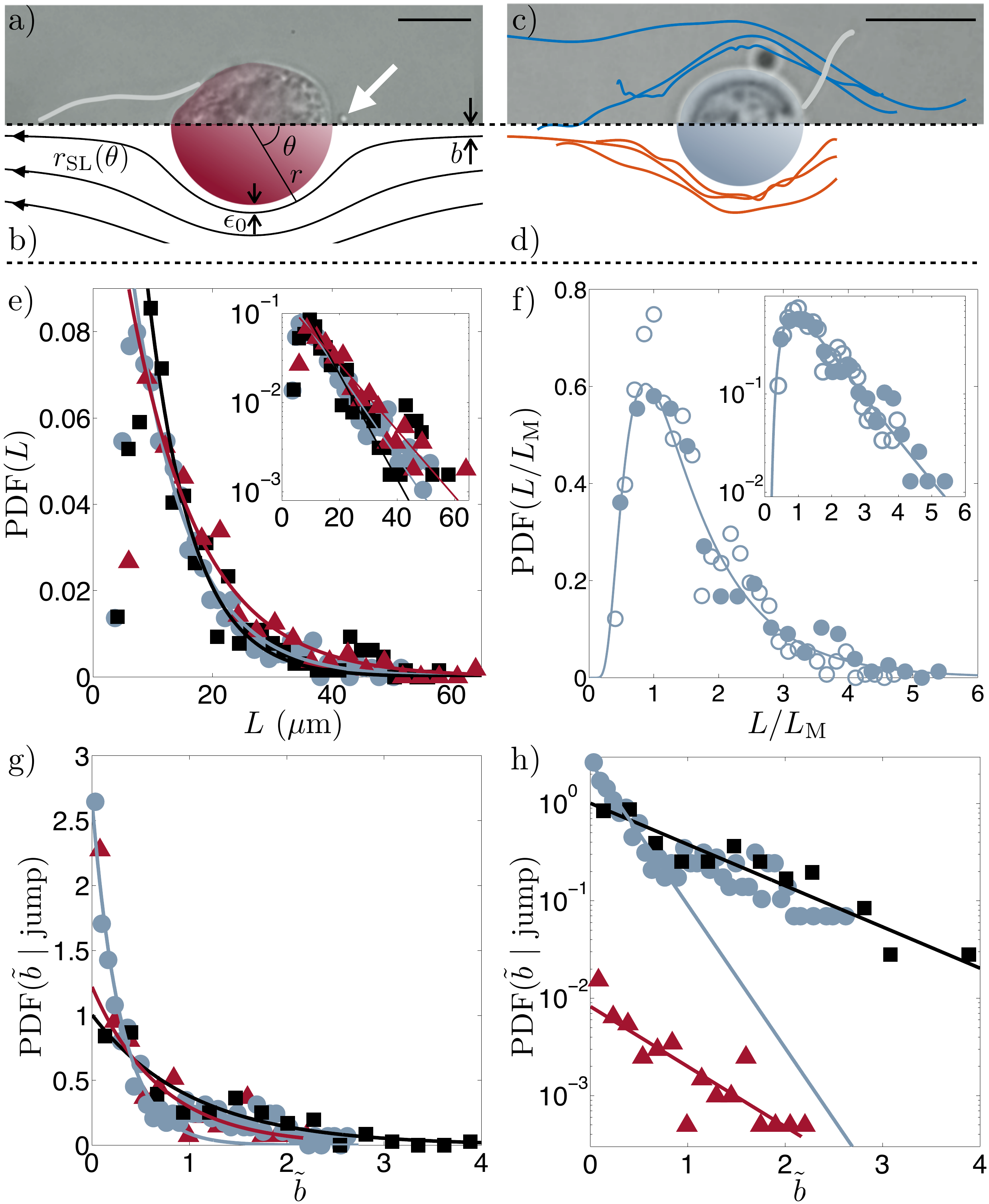}
\caption{
a) Snapshot of typical particle entrainment events for OM. The colloid ($\RT=0.5 \mummy$) is shown with a white arrow. Scale bar: $10 \mummy$. 
b) Diagram and approximated streamlines $r_{\rm SL}(\theta)=\RS+b\sqrt{2/3}/\sin(\theta)$ of the swimmer-generated flow in its co-moving frame. 
c) (resp. d)) Typical experimental (resp. numerical) tracer trajectories in the frame of CR. Scale bar: $10 \mummy$. 
e) PDF of entrainment length $L$ obtained with tracers of radius $\RT=0.5 \mummy$ for three different organisms: red triangles: OM; grey circles: CR; black squares: TS. 
Above a peak at value $L_{\rm M}^{\rm (S)}$, those \ar{ are well fitted by an exponential distribution (phenomenological) with characteristic} length scale $L_{\rm J}^{\rm (S)}$. \ar{ Inset: semi-log plot of the same data.}
f) Comparison of experimental (filled grey circles) and numerical (empty grey circles) jump lengths distributions using our outboard CR, for small impact parameters and normalised by $L_{\rm M}$, the length at the maximum: $L_{\rm M}^{\rm (sim)}\approx14.5 \mummy$ and $L_{\rm M}^{\rm (exp)}\approx9.4 \mummy$. \ar{ The best fit using Eq. \ref{eq:EntrLengthPDF} (solid line) agrees well with the experimental data. Inset: semi-log plot of the same data.}
g) Conditional PDF of the rescaled impact parameter $\tilde b=2b/w$ given an entrainment event ${\rm PDF}(\tilde b~|~{\rm jump})$. 
Curves are fitted by exponential distributions with characteristic lengths: $\tilde b^{\rm (TS)}=1.0\pm0.3$, $\tilde b^{\rm (CR)}=0.30\pm0.05$ and $\tilde b^{\rm (OM)}=0.70\pm0.25$. Same color code as in e).
h) Same data in a semi-log plot with the curve for OM (red triangles) shifted to highlight the two regimes in the distribution for CR (grey circles).
}
\label{figure1}
\end{figure}

\section{Nearby object-microswimmer interactions are governed by a universal entrainment mechanism}

When inspecting tracer dynamics at large magnification ($\times 100$) and high framerate ($500$ fps), very similar entrainment events are observed for all our organisms, regardless of their propulsion mechanism or generated flow.
Examples of typical trajectories with particle radius $\RT=0.5 \mummy$ are shown in Movies S1-3, for CR, TS and OM respectively \cite{Supplementary}. 
This mechanism is best understood from the viewpoint co-moving with the swimmer.
After an almost head-on collision, a bead reaches a region near the cell surface, Fig.~\ref{figure1}a).
It travels slowly around the body approximately following streamlines from front to back, and eventually leaves behind the organism, Fig.~\ref{figure1}b,c).
Since the only common physical feature of these organisms is the presence of the cell body surface, we propose that entrainment is only the consequence of the {\it no-slip} layer that this boundary induces.
Therefore, the particle is hydrodynamically coupled to the swimmer in this layer and resides in its vicinity for an extended duration, the contact time $T$.
In the laboratory frame, the particle is then displaced a distance $L$ in the direction of motion.
Hence, the average contact time and entrainment length are directly related via
\begin{align}
\label{eq:LengthTime}
\langle L \rangle  &\approx \VS \langle T \rangle .
\end{align}
To quantify our observations, we measure the distribution of $L$ for different swimmers, Fig.~\ref{figure1}e) \ar{ and \ref{figure1}e)-Inset}.
These \ar{ have all a similar shape, indicating a common underlying mechanism}, with \ar{an exponential-like} decay of length scale $L_{\rm J}^{\rm (S)}$ above the length $L_{\rm M}^{\rm (S)}$ at the peak of the distribution. 
Exponential fits to these curves give $L_{\rm J}^{\rm (CR)}=9.2\pm0.6 \mummy$ for CR, $L_{\rm J}^{\rm (TS)}=7.4\pm1.2 \mummy$ for TS and $L_{\rm J}^{\rm (OM)}=11.7\pm1.4 \mummy$ for OM, while we find $L_{\rm M}^{\rm (CR)}\approx6.7\mummy$, $L_{\rm M}^{\rm (TS)}\approx8.6\mummy$ and $L_{\rm M}^{\rm (OM)}\approx7.5\mummy$.

\ar{ Unexpectedly}, the characteristic entrainment length $L_{\rm J}^{\rm (S)}$ is significantly smaller for TS than CR despite a slightly larger body size. As described also in Appendix B, this effect is mainly attributed to the larger number of flagella in TS, which limit the average contact time by either rapidly pushing the beads backwards during the power stroke (Movie S2) or by ejecting them out of the no-slip layer during the recovery stroke. 
However, at the same time, front-mounted flagella can reach out and pull beads towards the body, widening the effective cross-section for entrainment around the swimming path. 
This is reflected in ${\rm PDF}(\tilde b~|~{\rm jump})$, the distribution of rescaled impact parameters, $\tilde b=2b/w$, before entrainment. For both TS and OM, ${\rm PDF}(\tilde b~|~{\rm jump})$ can be described accurately by a single exponential decay with characteristic lengths $\tilde b^{\rm (TS)}=1.0\pm0.3$ and $\tilde b^{\rm (OM)}=0.70\pm0.25$ (Fig.~\ref{figure1}g,h)). However, with CR, structurally very similar to TS but with a single pair of flagella, the distribution shows two markedly distinct behaviours for impact parameters above and below the cell body radius. For $\tilde b>1$, ${\rm PDF}(\tilde b~|~{\rm jump})$ follows the curve characteristic of TS; below that threshold we observe instead an exponential decay with a significantly smaller characteristic length $\tilde b^{\rm (CR)}=0.30\pm0.05$.  
This is a consequence of the fact that, in CR,  entrainments below $\tilde b\sim1$ are \ar{by and large a}  consequence of ``pure'' collisions with the cell body, without appreciable influence from flagella. Flagella participate instead in entrainment events with relatively large impact parameters, by increasing significantly their abundance over what would otherwise be expected. 
In order to study the main entrainment mechanism for CR, in the following we will focus on jump events with $\tilde b<0.75$ when comparing with numerical data. These include $\sim70\%$ of all the entrainments observed.

The outboard swimmer model captures the entrainment mechanism faithfully. 
Figures~\ref{figure1}c,d) show that simulated tracer trajectories reproduce well the experimental ones. In particular, we see that in both cases the particles tend to detach close to the swimming axis despite the variation in initial impact parameter, here randomly chosen in $[0,\RS]$ (see also Movies~S4,5 for CR and OM resp. and Movie S6 for a comparison \ar{with a model {\it Escherichia coli} (EC)}).
More quantitatively, the PDF of entrainment lengths from simulations agrees very well with the experimental one in Fig.~\ref{figure1}f),
\ar{when equivalent quantities are compared (i.e. the projection of the three-dimensional jump onto the focal plane)}.
Note that the entrainment length is globally overestimated ($L_{\rm M}^{\rm (sim)}\approx14.5 \mummy$ and $L_{\rm M}^{\rm (exp)}\approx9.4 \mummy$), \ar{which we} attribute to the approximations of this minimal model and the effect of flagella as discussed above.

\begin{figure}
\centering
    	\includegraphics[width= \linewidth]{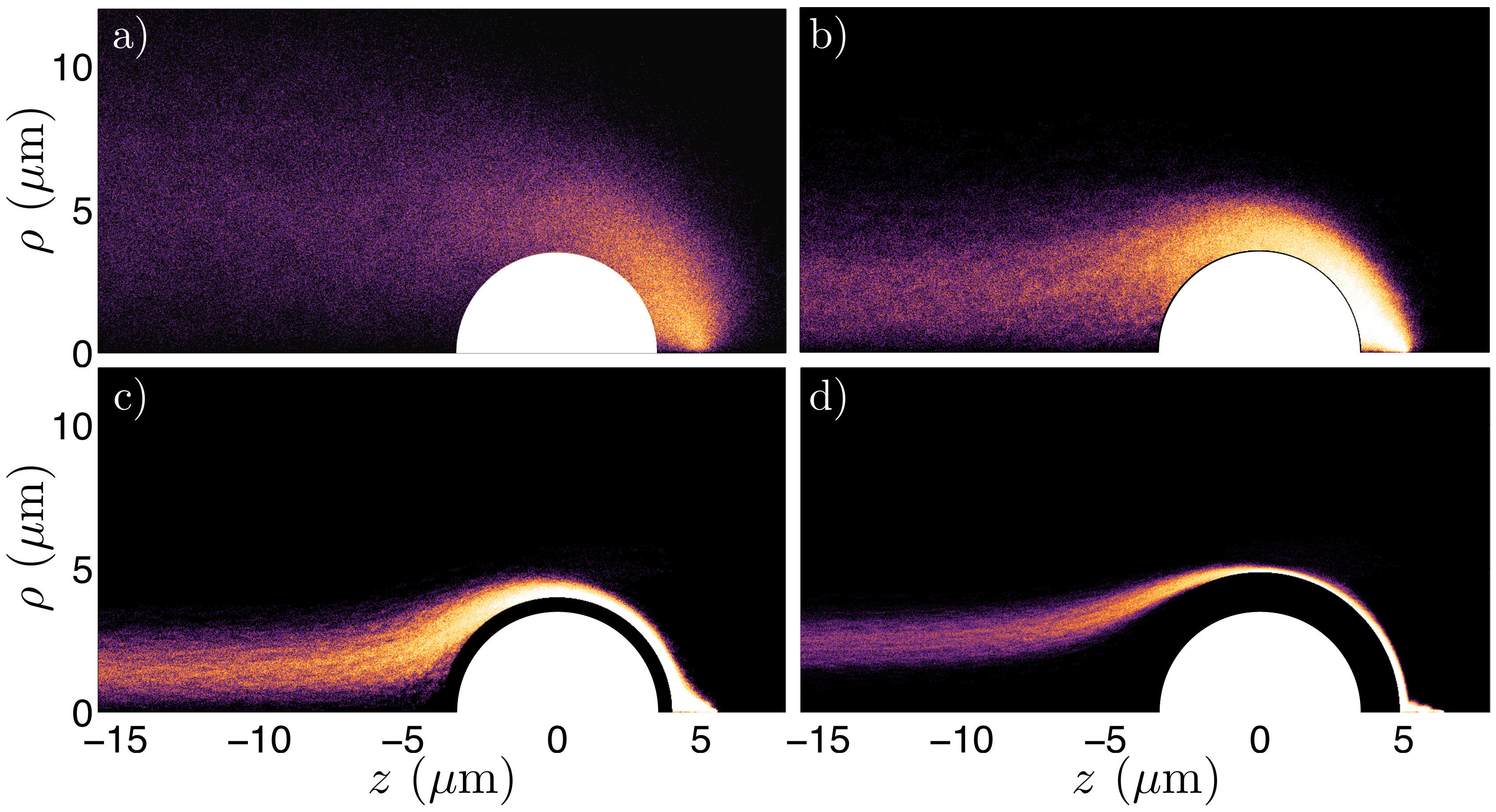}
\caption
{
Entrainment simulations by the outboard swimmer of Brownian tracer particles of various sizes. 
Shown are spatial PDFs of the beads, azimuthally and time-averaged, as seen in the co-moving frame of a CR alga, obtained by averaging over an ensemble of $10^3$ particles released in front of the body at $b=0 \mummy$ with a constant surface-to-surface distance of $1.5 \mummy$.
a) Small tracers with $\RT = 0.01 \mummy$ diffuse away quickly and are not entrained for a long time.
b-c) Intermediate-sized tracers with $\RT = 0.072, 0.52 \mummy$ primarily flow along streamlines close to the swimmer body, in its no-slip layer, and are the furthest entrained.
d) Large tracers with $\RT = 1.38 \mummy$ flow along paths far from the swimmer body, and are entrained less.
}
\label{figure2}
\end{figure}

\begin{figure*}
\centering
    	\includegraphics[width=0.97\linewidth]{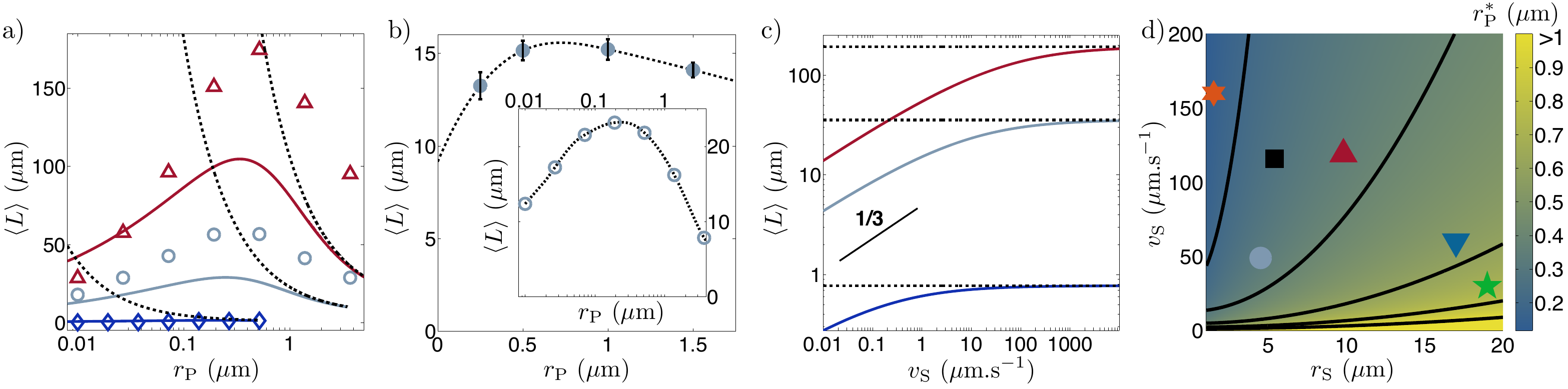}
\caption
{Average entrainment length $\langle L \rangle$ as a function of tracer size:
a) Simulated as in \fig{figure2} for OM, CR, and EC (open red triangles, grey circles and blue diamonds respectively).
The initial surface-to-surface distance is $11 \mummy$ for OM, $1.5 \mummy$ for CR, and $0.5 \mummy$ for EC. 
Solid lines show the corresponding analytical prediction; dashed lines without noise. Same data on log-log scale in Fig. S8b.
b) Average entrainment length obtained experimentally with CR for different tracer sizes. {\color{blue} The error bars represent the standard error on the mean (s.e.m.).}
Inset: Same data obtained in simulations with tracers initially located in front of the swimmer at impact parameters $b$ uniformly distributed in $[0, \RS]$. Dashed lines are guides to the eye. 
c) Analytical $\langle L \rangle$ as a function of swimming speed at fixed $\RT=1 \mummy$  (Solid lines; OM red, CR grey, EC blue. Dashed lines; without noise). 
d) The optimal bead size $\RT^*$ as a function of swimmer size and speed, obtained from equation \ref{eq:MeanTimeCubic} with constant value $\lambda=4$. Markers represen a few typical model organisms; grey circle: CR, black square: TS, red pyramid: OM, blue triangle: \textit{Euglena gracilis}, orange asterisk: \textit{Bdellovibrio bacteriovorus}, five-pointed green star: \textit{Peranema trichophorum}.
}
\label{figure3}
\end{figure*}

\section{Steric interactions reduce the contact time of large particles}

In light of these findings about the entrainment mechanism, we will now develop a simple theory to predict contact times. A detailed derivation is given in Appendix D.
We consider small particles that do not disturb the swimming direction significantly, which is a reasonable assumption if $\RT \lesssim \RS$ \cite{shum2017entrainment} as supported also by our experiments.
A streamline along the swimmer body can be approximated as
\begin{align}
r_{\textmds {SL}}(\theta,b) &\approx \RS+b\sqrt{2/3}/\sin \theta,
\end{align}
written in spherical coordinates of the reference frame co-moving with the swimmer, Fig. \ref{figure1}b).
Close to the cell body the advection of tracers is then governed by the tangential flow, $u_\theta(r, \theta)$.
Evaluating $u_{\theta}$ along a streamline with impact parameter $b$ gives the velocity of a particle,
\begin{align}
\label{eq:Tangential_1}
u_\theta [r_{\textmds {SL}}(b)]&=\frac{3 \VS \epsilon_0}{2 \RS} \frac{1}{g(\lambda)}
+ \mathcal{O}\left( \epsilon_0^2 \right),
\\
g(\lambda)&= \left(1+\frac{3\lambda^3(1+\lambda)^2}{(1+2\lambda)(1+\lambda^2)^{5/2}}\right)^{-1},
\end{align}
where $\epsilon_{0}=r_{\textmds {SL}}(\pi/2)-\RS=b\sqrt{2/3}$ is the closest distance of approach, Fig.~\ref{figure1}b), and $g(\lambda) \in [0,1]$ characterises the flagellar distance from the body ($\lambda=2,4,5$ for OM, CR and EC respectively, see Appendix C).
Then, in the deterministic limit, the contact time is found by integrating the inverse velocity along the particle trajectory, $T \approx \int ds/v_\theta$, where $ds$ is the arclength differential.
Since the velocity satisfies the no-slip condition at the cell surface, tracers with a small distance of approach $\epsilon_0$ move around the body slowly (\eq{eq:Tangential_1}) which increases the contact time. As the minimum distance is dictated by the finite size of the particle,  larger tracers are expected to experience stronger tangential flows and therefore smaller contact times.

\section{Brownian noise limits entrainment of small particles}

Together with particle advection, it is important to consider the presence of thermal noise. We first explore its effect by simulating outboard swimmers with tracers subjected to Brownian motion (Movies S4-6 \cite{Supplementary}).
Figure~\ref{figure2} shows how the spatial PDF of an ensemble of tracers initially in front of the microorganism depends on tracer size. 
Small tracers spread far  from the swimmer and do not efficiently access the no-slip layer (Fig.~\ref{figure2}a)).
Consequently, they are exposed to stronger tangential flows, limiting the contact time.
Large tracers do not diffuse away but cannot approach the no-slip surface closely, as described previously and illustrated by the inaccessible region around the cell body (Fig.~\ref{figure2}d)). 
Instead, we see a maximum in the contact time for beads of intermediate size, which concentrate most tightly around the moving cell (Fig.~\ref{figure2}b,c)). 
Optimal entrainment is systematically observed in simulations of both OM and CR, with optimal tracer radii $r_{\rm P}^*\sim 0.4,0.3 \mummy$ respectively (Fig.~\ref{figure3}a)).  Moreover, whereas these organisms can move particles along for $\sim 5$ body lengths, entrainments by the EC model do not exceed $\sim 1.2 \mummy$ for any tracer size.
The super-linear growth of the average jump length on cell size (Fig. \ref{figureSI2}c in Appendix D; see also \cite{mueller2017fluid}) strongly restricts particle transport for micron-size organisms, and is consistent with the lack of previous reports of strong entrainment by bacteria \cite{Wu2000, Ishikawa2010, Kasyap2014, Jepson2013}.

Experiments with CR for a range of different tracer sizes confirm the existence of a maximum in entrainment length (Fig.~\ref{figure3}b)), with diffusion-dominated tracer trajectories below, and steric-interaction-limited paths above, the optimal size $r_{\rm P}^* \sim 0.7 \mummy$.  These features are observed in simulations also when considering tracers that are initially located at random impact parameters $b$ within $[0,\RS]$, rather than directly in front of the cell (Fig.~\ref{figure3}b) inset). 
Altogether, the semi-quantitative agreement on both the location of the maximum and the values of jump lengths suggests that our outboard model captures the essential physics behind the entrainment mechanism.

\section{Optimal size for contact time}

\begin{figure}
\begin{center}
   	\includegraphics[width=\linewidth]{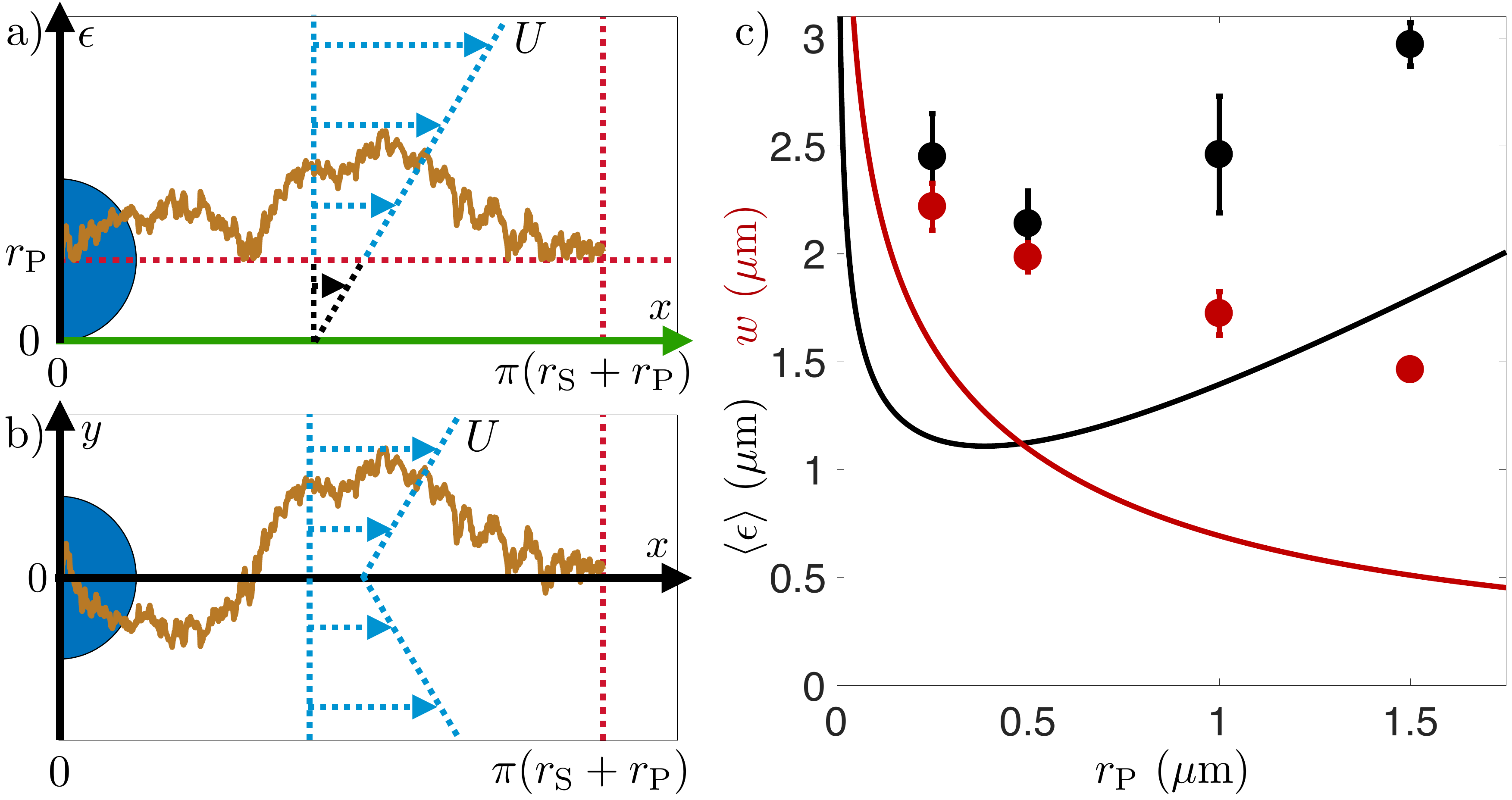}
\end{center}
\caption
{Schematics of the analytical model: 
a) A Brownian particle is advected over a solid surface (green axis) in a linear shear flow along the x-direction characterized by a strain rate $U$. The tracer of radius $\RT$, initially located at $x=0$ and $\epsilon=\RT$, is free to diffuse in any direction, but cannot cross the line $\epsilon = \RT$ due to steric interactions.
b) This situation is equivalent to a tracer free to diffuse through the surface $y=0$, but with a modified flow that is non-zero at the wall and always positive. \ar{ c) Parameters $\langle \epsilon\rangle$ (black circles) and $w$ (red circles) obtained from the fitting procedure of PDF$(L)$ for the four tracer sizes probed with CR (Eq. \ref{eq:effectiveCoefficients}). The analytical estimates of these parameters (full lines) are in good semi-quantitative agreement. In particular the minimum in $\langle \epsilon\rangle$ is captured at the right location.}
}
\label{figure4}
\end{figure}

Results from experiments and simulations can be rationalised with an approach akin to Taylor's dispersion \cite{Taylor1953b, stone2004engineering}.
Consider a Brownian particle advected in a linear shear flow over a straight solid surface that mimics the swimmer's cell wall, Fig.~\ref{figure4}a).
The flow velocity is $\vec{u} = \epsilon U  \vec{e}_x$, where the strain rate, $U=3 \VS /(2\RS g(\lambda))$, derives from the velocity along a streamline given by \eq{eq:Tangential_1}. 
A particle of radius $\RT$ is initially positioned at $(x=0,\epsilon=\RT)$, disperses with \ar{ thermal diffusivity  $D_0$} and is advected by the flow $\vec{u}(\epsilon)$, but cannot cross the line $\epsilon=\RT$. 
Without loss of generality, this is mapped to an unbounded ``image'' system \cite{vankampen1983} where the particle is initially located at $(x=0,y=0)$, the modified flow is $\vec{v} = (\RT + |y|)U \vec{e}_x$, and the tracer can diffuse everywhere (Fig.~\ref{figure4}b)).
\ar{Our first aim is to estimate} the average time $\langle T \rangle$ needed for the colloid to travel a distance $S = \pi(\RS+\RT)$ along the positive $x$-direction, imitating a journey around the swimmer's body. 
The \ar{ motion of the colloid is described by} 
\begin{align}
\label{eq:StochasticEOM1_1}
\dot{x}(t) &= \big(\RT + |y| \big)U + \xi_x(t); \quad
\dot{y}(t) = \xi_y(t),
\end{align}
where \ar{ $\vec{\xi}$ is a  Gaussian white noise satisfying $\langle \xi_i \rangle = 0$ and $\langle \xi_i (t) \xi_j (t')\rangle = 2D_0 \delta_{ij} \delta(t-t')$}. 
Integrating and ensemble averaging \eq{eq:StochasticEOM1_1} (\ar{see Appendix D}), leads to
\begin{align}
\label{eq:MeanPositionX2}
\langle x(t) \rangle  =  \RT U t + \ar{\frac{4}{3} \sqrt{\frac{D_0}{\pi}}} U t^{3/2},
\end{align}
\ar{ and requiring that $\langle x(\langle T\rangle) \rangle =S$, the mean contact time $\langle T \rangle$ becomes} the solution of the cubic equation
\begin{align}
\label{eq:MeanTimeCubic_1}
0 &= c_0 + c_2 \langle T \rangle + c_3 \langle T \rangle^{3/2},
\\
\label{eq:MeanTimeCubic2}
c_0 &= - \frac{2 \pi \RS (\RS+\RT) g(\lambda)}{3 \VS \RT} ;~
c_2 = 1;~
c_3 = \ar{\frac{4}{3\RT}\sqrt{\frac{D_0}{\pi}}}.
\end{align}
\ar{This can be solved analytically using Cardano's formula (Eq.~\ref{eq:Cardano} Appendix D) and then converted into the average entrainment length $\langle L \rangle$ with \eq{eq:LengthTime}. Figure~\ref{figure3}a) compares the results for CR, OM and EC (solid lines) when employing the same parameters used in the simulations, and shows that this simple approach recovers the correct qualitative non-monotonic behaviour of the average entrainment length, as well as  the position of the entrainment maxima.} 
The predicted magnitude of the entrainment length deviates by a factor of $\sim2$ near the maxima, due mostly to an overestimate of the average tangential speed experienced by the tracer (see Fig.~\ref{figureSI2}a and Appendix D4).
\ar{For either decreasing $D_0$ or increasing $\VS$, this simple estimate recovers the correct deterministic limit, in which $\langle L \rangle$ becomes independent of $\VS$ (\ar{Fig. \ref{figure3}c) dashed lines; see Appendix D2} and \cite{mueller2017fluid}).
However, as $\VS \to 0$ thermal noise becomes important, and $\langle L \rangle \sim \VS^{1/3}$(Fig.~\ref{figure3}c) solid line):
 slower organisms should display shorter entrainment lengths because particles diffuse away before being displaced substantially.
Altogether, these results suggest already} that Brownian motion can have \ar{ significant effects} on the entrainment efficiency.

%

\ar{A simple argument can also recover the optimal  tracer size $r_{\rm P}^*$ and its dependence on the swimmer's size and speed.}
Fig.~\ref{figure3}d) presents a map of $r_{\rm P}^*$, obtained by optimising the contact time (\eq{eq:MeanTimeCubic_1}) with varying $\RS$ and $\VS$ but fixed $\lambda=4$. 
On one hand, increasing swimmer speed $\VS$ shifts $r_{\rm P}^*$ to lower values, because faster advection limits the importance of Brownian motion.
On the other hand, increasing swimmer size $\RS$ increases $r_{\rm P}^*$, because the larger distance to travel around the cell enhances the relative effect of diffusion. 
A simple estimate for the optimal particle size is found by comparing a characteristic diffusion time \ar{$\tau_{\rm diff}\sim \RT^2/(2D_0)$} and a characteristic advection time $\tau_{\rm adv}\sim 2 \pi \RS^2/(3 \VS \RT)$ from \eq{eq:Tangential_1}. Defining the entrainment P\'eclet number as \ar{${\rm Pe}=3 \VS \RT^4/(4\pi \tilde D_0 \RS^2)$, where $\tilde D_0=\RT D_0$}, the optimal tracer size corresponds to  ${\rm Pe}=1$, which yields
\begin{align}
\label{eq:OptimalTracerSize}
r_{\rm P}^* &\simeq \sqrt[4]{4\pi \RS^2 \ar{\tilde{D}_0} / 3\VS}.
\end{align}
\ar{ Using experimental parameter values, this expression predicts $r_{\rm P}^* \sim 0.8 \mummy$ for CR, which compares well with the value $\sim0.7 \mummy$ found experimentally; and recovers the power laws describing the dependence of $r_{\rm P}^*$ on $\RS$ and $\VS$ observed in numerical solutions of \eq{eq:MeanTimeCubic_1} (Fig. \ref{figureSI3} in Appendix D).}

\ar{Notice that in many natural situations both parties can be active, and active motion of the small species could affect the duration and optimality of the contact process substantially. Considering as an example a system with a large predator and a small motile prey, the velocity $\VT$ and reorientation timescale $\tau_r$ of the latter lead to an effective diffusion coefficients $D_\textmds{active} \sim \VT^2 \tau_r$, which can easily be $\sim100$ times larger than $D_0$ for typical bacteria \cite{berg1993random, saragosti2012modeling}.
The modified entrainment P\'eclet number can be significantly smaller than the previous one, indicating a possibly substantial decrease in contact time. Therefore, even though prey motility increases encounter rates with predators \cite{Almeda2016}, it could nonetheless reduce the overall probability of being captured.}

\section{The contact time distribution}

\begin{figure*}[t!]
\begin{center}
 \includegraphics[width=0.9\textwidth]{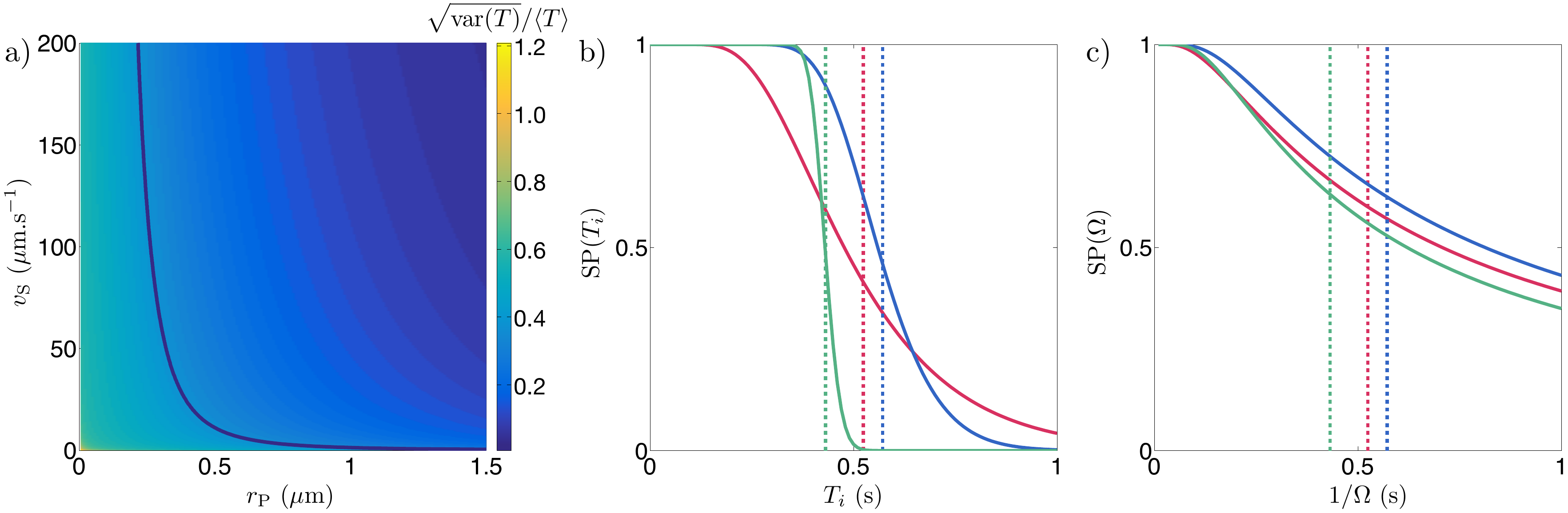}
\caption {
\ar{Contact time fluctuations can affect the probability of a successful entrainment interaction.
a) Map of the relative width $\sqrt{{\rm var}(T)}/\langle T\rangle$ of the contact time distribution as a function of tracer size $r_{\rm P}$ and swimming speed $v_{\rm S}$: 
Slow swimmers and small tracers show larger relative fluctuations in contact time. The solid line is the optimal tracer size for a given swimming velocity, \eq{eq:MeanTimeCubic}.
b) and c) Success probabilities of cell-object interactions for three different particle sizes ($r_{\rm P}=0.1 {\rm \mu m}$ - pink curve; $r_{\rm P}=0.7 {\rm \mu m}$ - blue curve; $r_{\rm P}=1.6 {\rm \mu m}$ - green curve) in two simple cases; In b) the interaction requires a finite time $T_i$ to happen, and in c) the interaction is described by a constant rate of success $\Omega$ per unit time. In the first case, fluctuations are important in setting the relative probabilities of success, while in the latter case the average contact time $\langle T\rangle$ (vertical dashed lines) mainly determines the chances of successful interaction.}}
\label{figure5}
\end{center}
\end{figure*}
%

\ar{The previous section provides a simple description of the average motion of a colloidal particle, based on its advection along the swimmer's body at an effective speed that depends on the particle's average distance from the swimmer Eq. \ref{eq:StochasticEOM1_1}. This 1D parallel can in fact be pushed further to describe the whole distribution of entrainment lengths ${\rm PDF}(L)$ in Fig.~\ref{figure1}e,f).  In the spirit of G. I. Taylor's work on diffusion within a pipe \cite{Taylor1953b}, the 2D motion of the entrained colloids in Fig.~\ref{figure4}a,b) (described by Eq.~\ref{eq:StochasticEOM1_1}) can be reduced to a 1D advection-diffusion process with an effective velocity $V_\text{eff}$ and diffusivity $ D_\text{eff}$ (see Appendix E). For this system, the distribution of first arrival times to a downstream boundary at a distance $S$ is well known, and can then be translated with Eq.~\ref{eq:LengthTime} into the entrainment length distribution: 
\begin{align}
\label{eq:EntrLengthPDF}
{\rm PDF}(L)&=\frac{S}{\sqrt{\frac{4\pi D_\text{eff}}{v_{\rm S}} L^3}}\exp{\left(-\frac{v_{\rm S} \left( S-\frac{V_\text{eff}}{v_{\rm S}}L \right)^2}{4 D_\text{eff} L}\right)}.
\end{align}
This functional form provides an excellent fit to {\it all} the experimental distributions recorded (see \fig{figure1}f) and Fig. \ref{figure10}a,b) in Appendix E), confirming that the main entrainment behaviour can indeed be studied within this simplified framework.
Simple analytical estimates for $V_\text{eff}$ and $D_\text{eff}$ can then be found by approximating the 2D advection-diffusion process as  being confined within a uniformly sheared region of appropriate thickness (Appendix E):
\begin{align}
\label{eq:effectiveCoefficients}
V_{\rm eff} = U \langle \epsilon\rangle; &\quad D_{\rm eff} = D_0\left(1+ \frac{U^2 w^4}{120D_0^2}\right),
\end{align}
where $\langle \epsilon\rangle = \RT + \sqrt{4D_0\langle T\rangle/\pi}$  and $w= (\langle\epsilon\rangle-\RT)(1{\color{blue}+}\sqrt{\pi/2-1})$ are the average position and estimated spread of the particle distribution above the swimmer's surface at time $\langle T\rangle$.
The comparison between the fitted and calculated values of $\langle \epsilon\rangle$ and $w$ for the different tracer sizes probed with CR in Fig.~\ref{figure4}c) shows that this simple description captures well the qualitative dependence of the experimental entrainment length distributions on tracer size; and provides an accurate prediction for $\RT^*$, which corresponds to the minimum of $\langle \epsilon\rangle$. At the same time, Fig.~\ref{figure4}c) shows that  the parameter $\langle \epsilon\rangle$ appears to be globally underestimated by $\sim1\,\mu$m. This is likely the result of the approximations involved in the derivation of Eq.~\ref{eq:effectiveCoefficients} rather than a consequence of factors like electrostatic interactions (the Debye screening length in the algal media is easily estimated to be always $\lesssim 5\,$nm).
}

\ar{An experimentally validated model of the entrainment process allows us to rationalise experimental features of the measured distributions, and explore  qualitatively their dependence on system parameters in lieu of labour-intensive experiments. On one hand, it is easy to see that the part of the distribution in Eq.~\ref{eq:EntrLengthPDF} past the maximum can indeed be described well by a single exponential decay if the parameters provide a sufficiently large value of the effective P\'eclet number ${\rm Pe}_{\rm eff}=SV_\text{eff}/D_\text{eff}$. This is what we observe in Fig.~\ref{figure1}e,f). 
}

\ar{ The same quantity also controls the effect of fluctuations, represented by the relative spread of the distribution of entrainment lengths, $\sqrt{{\rm Var}(L)}/\langle L\rangle\sim (1/{\rm Pe}_{\rm eff})^{1/2}$. This implies that slower cells will feature a wider distribution of entrainment lengths, and therefore contact time, as the effective diffusion plays a relatively larger role than in fast-moving cells (see Fig. \ref{figure5}a)). Similarly, small tracers feature larger deviation relative to their average contact time due to the dominant effect of the effective diffusion.
}

\ar{In turn, these fluctuations can have some effect on the likelihood of interactions that take place during entrainment. From the contact time distribution we can analytically extract the probability of success for a given cell-object interaction in two simple illustrating examples: 
i) the case where the interaction requires a minimum time of contact $T_{i}$ and 
ii) the case of a constant success rate $\Omega$ per unit time. 
In the first case, the success probability (SP) is given by
\begin{align}
 {\rm SP}(T_i)&={\rm Prob}(T>T_i) \\
 &=\frac{1}{2} \Big[1 + \text{erf}\left(\frac{S-T_i V_\text{eff}}{2 \sqrt{D_\text{eff} T_i}}\right)
\nonumber \\
& \quad \quad \quad  - e^{\frac{S V_\text{eff}}{D_\text{eff}}} \text{erfc}\left(\frac{S+T_i V_\text{eff}}{2 \sqrt{D_\text{eff} T_i}}\right) \Big],
\end{align}
which is shown Fig. \ref{figure5}b) for three different tracer sizes using the experimental CR parameters.
For large particles the contact time does not deviate much from its average value (${\rm Pe}_{\rm eff} \gg 1$), and the SP follows a switch-like dependence on the interaction time $T_i$ (green curve), where the switching time is the average contact time $\langle T\rangle$ (dashed lines). 
However, as the particle size decreases below the optimal size (i.e. $\RT < \RT^*$), the fluctuations \textit{enhance} the chance of success for slow interactions (i.e. $T_i>\langle T\rangle$), and small particles (red curve) can be more successful than those with the largest average contact time (blue curve) at slow interactions, but at the cost of reduced success for faster interactions.
In the second case, the success probability is
\begin{align}
 {\rm SP}(\Omega) &= \int_0^{\infty}\left(1-e^{-\Omega T}\right){\rm PDF}(T)dT \\
  &=1-\exp\left({\frac{S(V_{\rm eff}-\sqrt{V_{\rm eff}^2+4D_{\rm eff}\Omega})}{2D_{\rm eff}}}\right),
\end{align}
shown in Fig. \ref{figure5}c) for the same tracer sizes. Here we observe that fluctuations do not play a significant role. Regardless of the rate of success $\Omega$, the chance of success will always be optimal for the tracer with the largest average contact time (\eq{eq:MeanTimeCubic_1}). 
Despite their simplicity, these examples already portray the interesting role that the noisy entrainment process can play in different types of natural interactions.
}

\section{Conclusion}

Swimming microorganisms vary greatly both in body size and in the details of their propulsion, from the number and arrangement of flagella to their gaits. 
Yet, despite this variability, our results show that particle entrainment is a remarkably universal mechanism. 
Combining experiments with numerical simulations, we see that pullers and pushers entrain particles with similar efficiency. 
We see no evidence for either ``wake bubble'' effects \cite{Lin2011}, or entrainment due to a stagnation point in front of the cell \cite{Jeanneret2016, Thiffeault2015}.
Instead, our results suggest that entrainment is a consequence of an organism's {\it no-slip} surface, a characteristic shared by the three species we study here.
This feature, recently suggested also in \cite{mueller2017fluid}, is consistent with the lack of entrainment in numerical studies involving squirmers, which instead propel with a surface slip velocity \cite{Lin2011, Thiffeault2015}. Accordingly, we predict that ciliates like \textit{Paramecium} and multicellular algae like {\it Volvox} will not substantially entrain micron-sized objects, since they swim by an effective surface slip generated by thousands of cilia and flagella. Studies of {\it V. carteri} swimming through a colloidal suspension support this hypothesis (see supplementary movie from \cite{Drescher2010f}).
Comparing different species, we also see that the flagellar arrangement has a quantitative effect on particle entrainment. Front-mounted flagella decrease the average contact time $T$ but increase the interaction range, and are therefore likely to increase the frequency of entrainments.

The outboard model proves to be in fair qualitative and quantitative agreement with our experimental results. 
It provides directly comparable tracer dynamics and, crucially, \ar{it reproduces successfully} the shape of the entrainment length distribution. This strongly suggests that the model \ar{ captures correctly} the essential physics, with further support provided by the maximum in the entrainment length over particle size observed in both experiments and simulations. 
\ar{ These results can in fact be accurately described by a simple Taylor-dispersion theory, which provides the correct functional form for the entrainment distribution with parameters that can be estimated semi-quantitatively through simple approximations. The theory, which is based exclusively on Brownian diffusion within the high-shear layer close to the swimmer's surface, provides an intuitive justification for the existence of an optimal particle size for entrainment, $\RT^*$, set by the balance between diffusive and advective timescales.}

Size-dependent contact times might affect predation by microorganisms. 
Experimental studies of microbial grazing indicate that this is indeed a selective process \cite{Montagnes2008a}.
For example, \textit{Oxyrrhis marina} feeds on prey ranging from bacteria to cells as large as itself \cite{flynn1996prey, roberts2010feeding}, but seems to have an optimal prey size \cite{hansen1992prey, hansen1996grazing, roberts2010feeding, davidson2011oxyrrhis}, in agreement with our hydrodynamic arguments. 
Phagotrophic selectivity is complex, \ar{and surprisingly common even amongst microbial species historically considered exclusive autotrophs (e.g. some green microalgae \cite{selosse16}). It} certainly depends on many factors including chemical cues and cell surface properties.
However, the physics leading to the non-monotonic size-dependence of contact time is inescapable, and therefore needs to be taken into account.
A non-monotonic dependence on tracer size has also been reported for the effective diffusion of colloidal particles suspended within an {\it E. coli} culture \cite{patteson2016particle}. These experiments, which focus on particles larger than the microorganisms, show that potentially new mechanisms could be at play in that size range.

To conclude, we have seen that particle entrainment is a generic feature of the interaction between microorganisms and small particles, and have characterised the physics behind it. 
A complete picture of these interactions, however, will require to integrate our results not only with those from intermediate and far-field studies \cite{Pushkin2013, mathijssen2015tracer}, but also with a thorough characterisation of the navigational strategy of the microorganism \cite{mendez2013stochastic}. 
This will enable more accurate bottom-up models of microbial grazing, which can be used to predict feeding or clearance rates by phagotrophs \cite{dolger2017swimming}, and potential trade-offs between feeding and swimming \cite{strathmann2006good, Michelin2011, gilpin2016vortex}. 
At the same time, prolonged contact also underpins the successful binding of viruses and other parasites to cells \cite{Seisenberger2001}. 
By showing that the contact time with motile microorganisms is limited, $\lesssim 3$~s for reasonable swimmer sizes and speeds, we suggest that motility can potentially affect infection rates \cite{Taylor2013} and thus provide a fitness advantage. This should be true in particular for ciliates, which display an effective surface slip and therefore a faster clearance of particles. Targeted experiments and modelling efforts in this area will improve our mechanistic understanding of early infection events in microorganisms. 

From a micro-engineering perspective, our model shows that entrainment lengths become millimetric or even larger for micrometric tracers when considering swimmers or active particles with radius $\RS \gtrsim 50 \mummy$. When combined with externally triggered reorientation events, this purely hydrodynamic phenomenon could enable cargo transport by self-propelled colloids without requiring any surface functionalisation. 

\begin{acknowledgments} 
We are grateful to Kirsty Wan and Raymond Goldstein for sharing the initial culture of TS, \ar{and to an anonymous Referee for encouraging us to search for an analytic expression of the distribution of entrainment lengths}.
This work was supported in part by and ERC Advanced Grant (291234 MiCE) \ar{and the Human Frontier Science Program (Fellowship LT001670/2017)} (A.M.); and a Royal Society Research Grant (RG150421) (M.P.).
\end{acknowledgments}

\appendix
\section{Experimental methods}
\label{sec:SIexperimentalMethods}

\begin{figure}
\centering
	\includegraphics[width=0.8\linewidth]{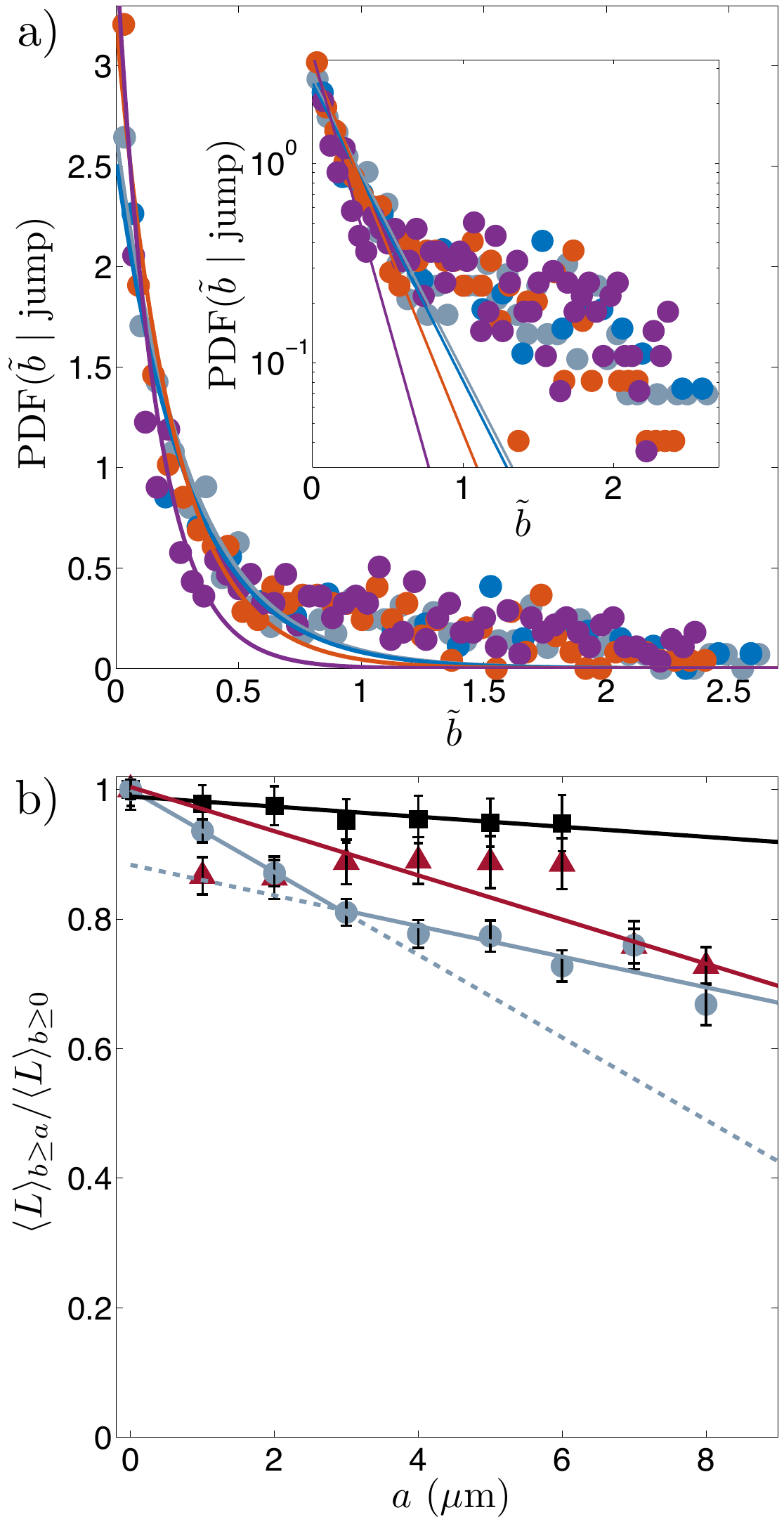}
  	\caption{(a) PDF of impact parameter $b$ prior a jump with CR for all tracer sizes explored. All curves present a bump above $\tilde b \approx 1$ showing the effect of flagella bringing the tracers toward the body of the cell. Color code: blue $\RT=0.25 \mu $m, grey $\RT=0.5 \mu $m, orange $\RT=1.0 \mu $m, purple $\RT=1.5 \mu $m. Inset: Same as main curve in semi-log plot to emphasize the bumps. (b) Entrainment length averaged over impact parameters $b$ larger than a given value $a$ showing the effect of flagella for both TS and CR. Same color code as in main text. 
}
\label{figureSI}
\end{figure}

\subsection{Introduction to the microorganisms}

We have considered 3 different species of flagellated eukaryotic unicellular microorganisms presenting different features of swimming but having roughly the same body-size: {\it Chlamydomonas reinhardtii} (CR), {\it Tetraselmis subcordiforms} (TS) and {\it Oxyrrhis marina} (OM). CR and TS are both green algae with a very similar prolate body shape but differing in the number of flagella in front of their body: 2 for CR and 4 for TS. CR uses most of the time a breaststroke way of swimming \cite{Polin2009a}, while TS flagella beat successively in a transverse gallop fashion \cite{Wan2016}.  They are puller-like microorganisms  \cite{Lauga2009,Guasto2010, Klindt2015}. The length and width of their body are respectively $\langle l^{\rm (CR)}\rangle=10.1\pm1.7 \mu$m and $\langle w^{\rm (CR)}\rangle=8.0\pm1.7 \mu$m for CR, $\langle l^{\rm (TS)}\rangle=13.7\pm1.9\mu$m and $\langle w^{\rm (TS)}\rangle=8.4\pm1.0 \mu$m for TS. In the (confined) microfluidic channels (thickness $h=25.8\pm0.1 \mu$m) used for the experiments, the average speed of these microorganisms were: $\langle v^{\rm (CR)}\rangle=49.1\pm2.5\mu$m.s$^{-1}$ and $\langle v^{\rm (TS)}\rangle=116\pm6\mu$m.s$^{-1}$. OM is a dinoflagellate widely distributed across the seas (except the polar areas). During the past twenty years it has become a model organism for studying predator-prey interactions at the micro-scale. It also has a prolate shape but is bigger and more asymmetric than CR and TS. We measured an average length $\langle l^{\rm (OM)}\rangle=22.9\pm3.4 \mu$m, an average width $\langle w^{\rm (OM)}\rangle=16.6\pm1.8 \mu$m and an average speed (in a $29 \mu$m-thick channel) $\langle v^{\rm (OM)}\rangle=119\pm10 \mu$m.s$^{-1}$. This microorganism has 2 flagella at the back of its body a long and a short one. The long one is used to propel while the short transverse one is used to turn and appears to be a tool to catch and recognize preys. We chose this microorganism as a representative of eukaryotic pusher-type swimmers \cite{Kiorboe2014}. 

\subsection{Cultures of the microrganisms}

Cultures of CR strain CC125 were grown axenically in a Tris-Acetate-Phosphate medium at $21^{\circ}\,$C under periodic fluorescent illumination ($100\,{\rm \mu E/m^2s}$, OSRAM Fluora) with a dark/light cycle of 12h/12h. This is done to synchronize the cell cycles among the population. Cultures were kept in the exponential phase (concentration of $\sim 5.10^{6}\,$cells/ml) by transferring daily the algae into a new flask of fresh-medium. 

Cultures of TS strain CCAP 161/1A were grown axenically in a Seawater Nutrient Broth medium at $21^{\circ}\,$C under fluorescent illumination ($100\,{\rm \mu E/m^2s}$, OSRAM Fluora). Cultures were kept at a concentration of $\sim 10^{6}$ cells/ml by transferring biweekly the algae into a new flask of fresh-medium (growth rate $\sim 1.4$ cells/day). 

Cultures of OM strain CCAP 1133/5 were grown monoxenically in a f/2 medium at $21^{\circ}\,$C together with the small alga \textit{Nannochloropsis oculata} (CCAP 849/1) serving as food. Fresh medium and algae were supplied to the culture every $\sim 1-2$ months. 

When performing the experiments, cultures were harvested in the exponential phase ($\sim 10^6$ cells/ml for CR and TS, $\sim 10^5$ cells/ml  for OM).

\subsection{Experimental setup}

After gently centrifuging the suspension of a given organism, the supernatant was replaced by the appropriate fresh medium also containing a small fraction of colloids (Polysciences, catalog no. 19819-1) of the required radius ($\RT=0.50\pm0.01 \mu$m for TS and OM, $0.26\pm0.005\leq \RT\leq 1.55\pm0.03 \mu$m for CR) at a concentration $\lesssim 10^{-4}\%$ solids. These polystyrene particles present carboxyl groups on their surface which contribute to prevent  adhesion to the microorganisms. The suspension was then loaded into a PDMS based microfluidic chip having a visualisation chamber $2$ mm wide and $\sim 26 \mu$m (for CR and TS) or $\sim 29 \mu$m thick (for OM).  Given the size of the organisms and the thickness of the chambers, there was enough room for the particles to travel over or beneath the swimmers around their bodies, making the entrainment mechanism fully 3D. This justifies the 3D numerical approach as the near-field flows are not expected to be influenced by the presence of confining walls. Because the colloidal suspension was very diluted and the depth of focus thick enough, we were still able to track the colloids before and after entrainment and extract accurately the entrainment lengths. The channels were previously passivated with $0.15\%$w/w BSA solution in water. The inlets of the chips were sealed with Vaseline to prevent evaporation. 

The systems were recorded at $25$ fps using a Pike camera (F-100B, AVT) under phase contrast illumination on a Nikon TE2000-U inverted microscope. A long-pass filter (cutoff wavelength 765 nm) was added to the optical path to prevent phototactic response of CR and TS. \ar{The magnification was set according to the size of colloids: $30\times$ for tracer radii of $0.5$ and $1$ $\mu$m, $40\times$ for $0.25 \mu$m tracers and $20\times$ for $1.5 \mu$m tracers. We limited the experimental investigation of the influence of tracer sizes on the entrainment to this size range because: i) visualisation of  tracers smaller than $\sim 0.2 {\rm \mu m}$ requires fluorescence, which influences the behaviour of the algae; ii) tracers larger than $\gtrsim 2 {\rm \mu m}$ influence the motion of the microorganisms, probably due to mechanosensation. This is an interesting range to explore but outside the scope of the present study. Despite these constraints,  the data in both Fig.~3b and~\ref{figure4}c display clearly the non-monotonic dependence of entrainment on particle size, with an optimum at a radius comparable to the predictions of both outboard model and the theoretical model.}

Organisms and colloids trajectories were then digitised using a standard Matlab particle tracking algorithm (The code can be downloaded at http://people.umass.edu/kilfoil/downloads.html). Individual jumps were extracted from the trajectories with the same procedure as our previous work \cite{Jeanneret2016}, \ar{ which was complemented by a visual inspection of every single events in order to filter out non-entrainment perturbations. This extra step was not present in \cite{Jeanneret2016}. For the experiments with CR, we have extracted the following number of jump events for each particle size: 388 for $r_{\rm P}=1.5 {\rm \mu m}$, 303 for $r_{\rm P}=1 {\rm \mu m}$, 311 for $r_{\rm P}=0.5 {\rm \mu m}$ and 135 for $r_{\rm P}=0.25 {\rm \mu m}$.}

\section{On the effect of flagella}

To quantify the role of the flagella on the entrainment process, we first consider the impact parameter $b$ \ar{preceding the entrainment}. It has to be noted here that in the experiments we can only measure the projected impact parameter on the focal plane due to the lack of vertical resolution for both the swimmers and the colloids. However, as will be clear in what follows, this measurement allows to extract insightful information. The PDF of impact parameters $b$ of organism-tracer encounters conditioned to the fact that the beads will be entrained ${\rm PDF}(b~|~{\rm jump})$ is shown Fig.~1g-main text for the three organisms after rescaling by the half-width of the organisms $w^{\rm (S)}/2$ (see Fig.~1h-main text for a semi-log plot). These PDFs are \ar{similar} for the three swimmers and can be well fitted by exponential distributions. However, the characteristic decay $\tilde b^{\rm (S)}=2b^{\rm (S)}/w^{\rm (S)}$ obtained for TS is larger than that of CR and OM: $\tilde b^{\rm (TS)}=1.0\pm0.3$ while $\tilde b^{\rm (CR)}=0.30\pm0.05$ and $\tilde b^{\rm (OM)}=0.70\pm0.25$, showing that TS can entrain particles even when those are relatively far away from the swimming path. We interpret this results as a consequence of the transverse gallop beating pattern of the four flagella of TS, \ar{increasing the probability} of the beads being brought by these appendages towards the body, whatever the impact parameter, small or large. This is not true for OM and CR for which we observe more peaked distributions around $\tilde b=0$, showing that particles \ar{are much more likely to be entrained} if close to the swimming path. However, the distribution for CR presents a clearly visible bump above $\tilde b\sim 1$, also observed with other tracer sizes, Fig.~\ref{figureSI}a. This increase in the probability of entrainment at larger $\tilde b$ is also explained by the presence of flagella that bring the beads towards the no-slip layer of the cell. \ar{This effect is much less important for CR because it only has two flagella}. 

Finally, we expect the entrainment length to decrease with impact parameter, because the larger the impact parameter $b$ the larger the parameter $\epsilon_0$ (Fig.~1b-main text) and consequently the further from the no-slip surface the bead will travel. To probe for this effect, \ar{we plot in Fig.~\ref{figureSI}b the entrainment length $\langle L\rangle_{b\geq a}$: the average displacement for impact parameters $b\geq a$ vs. $a$.} This reduces considerably the noise compared to simply looking at the average length at any given impact parameter. This quantity depends very weakly on $a$ for the quadriflagellate TS, showing that the entrainment length is a weak function of impact parameters. Again this is due to the flagella bringing the beads towards the no-slip surface but in an unpredictable manner: the beads reach the no-slip surface at {\it random} polar angles $\theta$ (Fig.~1b-main text) whatever the impact parameter $b$. For the biflagellate CR, we observe an interesting behavior where this quantity first decreases substantially up to $a\sim3-4~\mummy\sim w^{(CR)}/2$ while above the curve flattens. At small impact parameters the flagella do not play a role and the beads arrive at the surface of the cell in a more predictable way following relatively well defined pathlines. However at larger impact parameters, the flagella have an effect akin to TS and \ar{render the entrainment length more uniform}. For the pusher OM we seem to observe a decreasing curve with a constant slope, consistent with our interpretation of the role of the flagella with CR and TS. However the more unfrequent entrainment events for this organism limit substantially the statistics.

\section{\ar{Outboard model for entrainment}}
\label{sec:OutboardModel}

\begin{figure}[t]
    	\includegraphics[width= 0.8\linewidth]{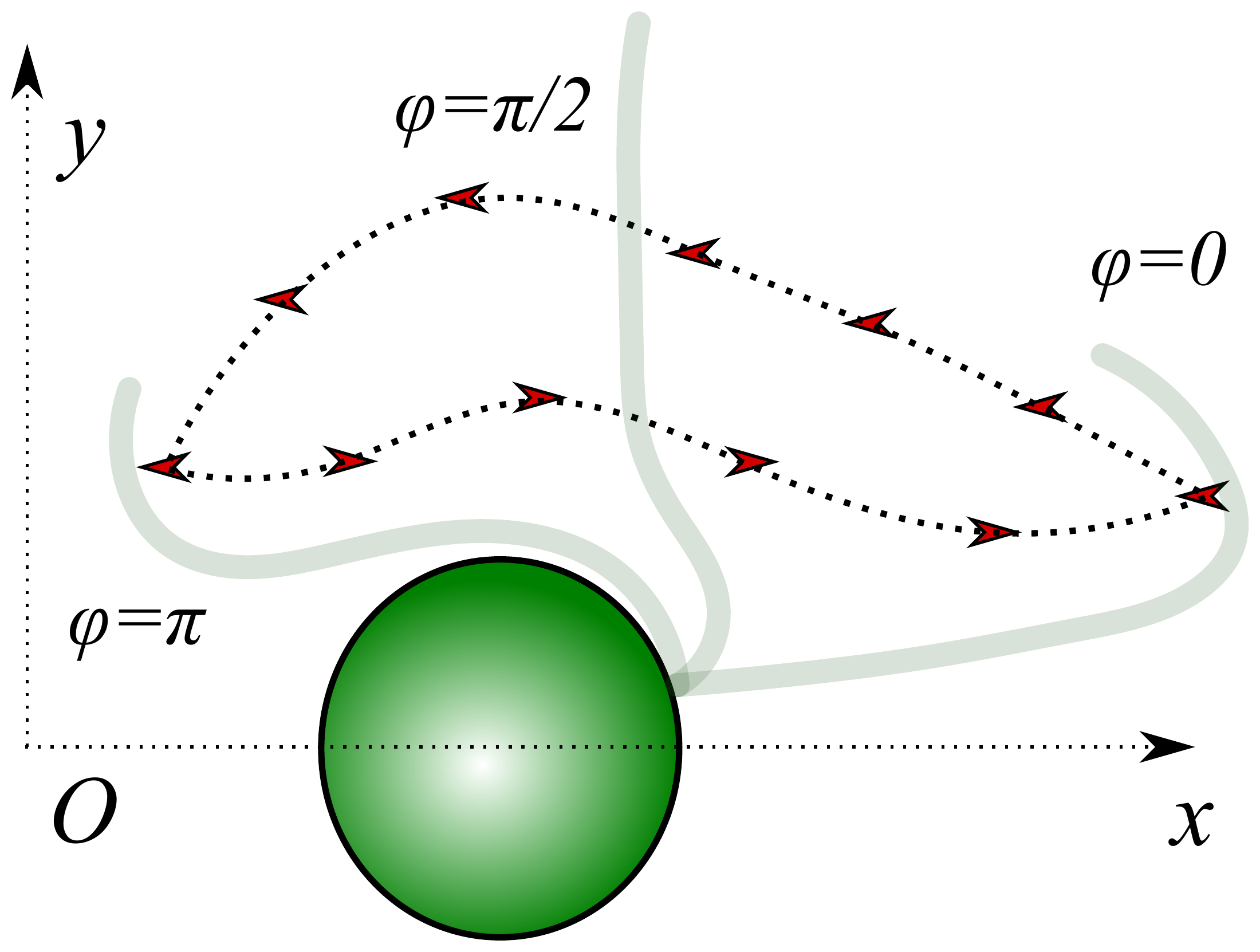}
  	\caption{
Diagram of the model \chlamy swimmer. The cell body is a sphere of radius $\RS$ (in green) that moves along the $x$ axis with velocity $\vec{v}_\textmds{S}(\phi)$, 
and the flagella are represented by two point forces (red arrows) that follow a loop-like trajectory $\vec{x}_\textmds{F}(\phi)$ around the cell body during the beat cycle (sketched in faint green for $\phi = 0,\pi/2,\pi$). 
}
\label{fig:Diagram}
\end{figure}

Here we present the `outboard propulsion' model to evaluate the flow fields generated by a micro-swimmer, which we use later to perform simulations with tracer particles. 
This name implies that all propulsion forces are transmitted to the liquid from outside the cell body, as opposed to be generated at the swimmer surface.
For example, the helical flagellum of a bacterium bears resemblance to an outboard motor, while CR takes after a rowing boat.
To capture the near-field flows of an organism, we use a finite-sized spherical body (radius $\RS$) with a \textit{no-slip} boundary condition at its surface.  
Propulsion is achieved by a set of regularised Stokeslets outside the body, whose flow satisfies the no-slip condition on the cell body. 
The number, arrangement and motion of the driving forces is species specific. 
Instantaneous swimming speed and rotation derive from the requirement of zero net force and torque. 
This model will be used to simulate an actively beating CR, a steady OM, and a steady \textit{E. coli} (EC) bacterium for comparison.

\ar{Our approach is inspired by previous work \cite{bennett2013emergent, friedrich2012flagellar, brumley2014flagellar, dolger2017swimming}, and is similar to the one developed at the same time by Mueller and Thiffeault \cite{mueller2017fluid}, who implemented a two time-dependent point-force model with a finite-sized no-slip body to simulate \textit{Chlamydomonas} cells.}

\begin{figure*}
	\includegraphics[width= 0.8\linewidth]{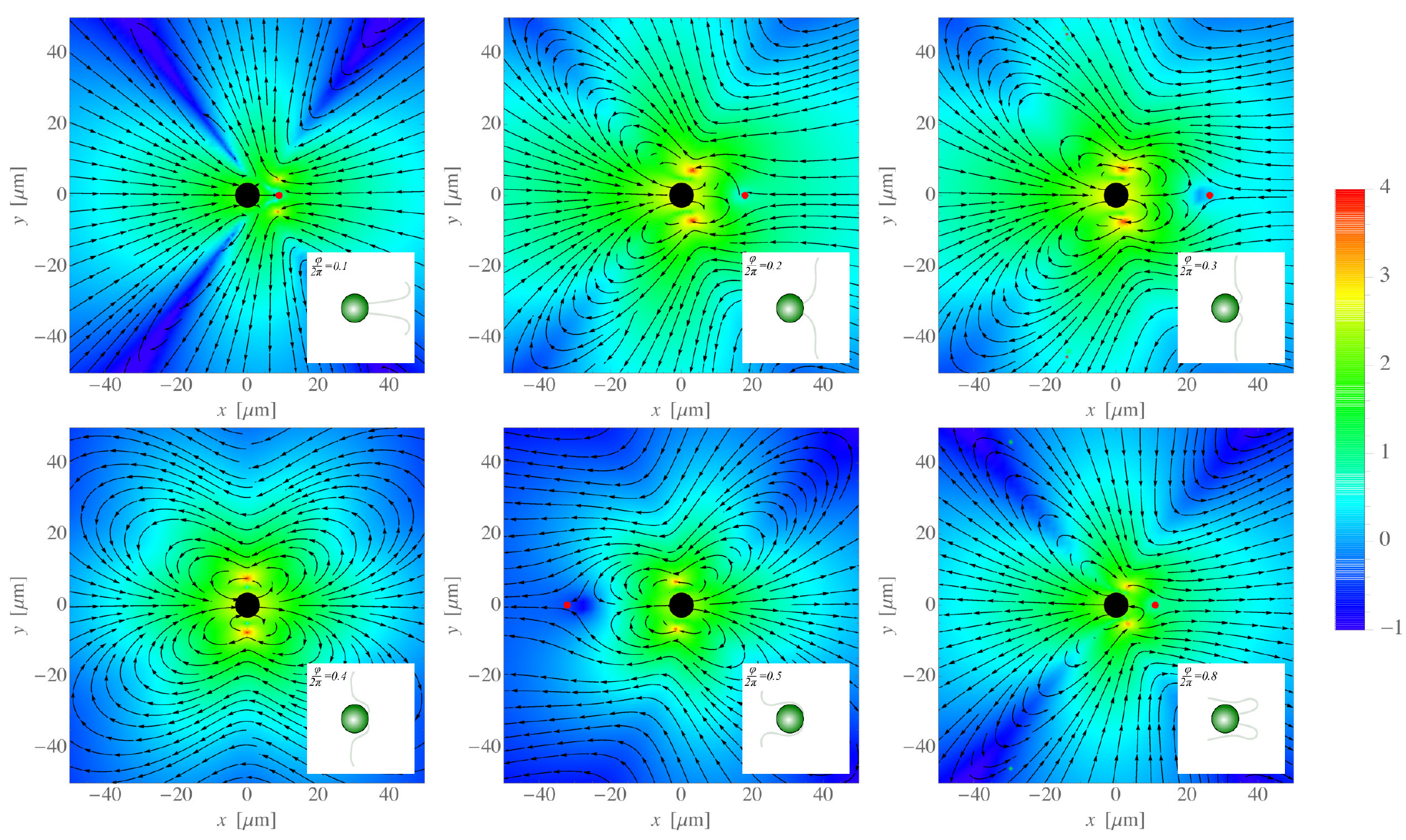}
  	\caption{
Flow fields generated by the model \chlamy, in the $(x,y)$ plane and in the lab frame, at six different times during the swimming stroke cycle ($10 \phi/2\pi = 1,2,3,4,5,8$).
The swimmer body (black disc) is oriented in the $x$ direction, and insets show sketches of the flagellar shape at each instance.
The velocity directions are shown by streamlines (black arrows) and the magnitude by colours, ranging from $10^{-1} \mummys$ (blue) to $10^{4} \mummys$ (red) on a logarithmic scale.
The analytically approximated position (\eq{eq:StagnationPointPosition}) of the stagnation point on the $x$ axis is indicated with a red point.
}
\label{fig:FlowFields}
\end{figure*}

\subsection{Model for \textit{Chlamydomonas}}
\label{sec:ModelChlamy}

We first consider \ar{CR} with a body that satisfies the no-slip boundary condition at its surface, and two beating flagella (see \fig{fig:Diagram}).
The cell body is modelled as a solid sphere of radius $\RS$ that is located at position $\vec{x}_\textmds{S}(t) = (x_\textmds{S},0,0)$ at time $t$ and oriented in the $x$ direction, in Cartesian coordinates.
The flagella are represented by two point forces (Stokeslets) that move along the trajectories $\vec{x}_\textmds{F1}(\phi)$ and $\vec{x}_\textmds{F2}(\phi)$ in the $(x,y)$ plane, where both flagella exert an equal force $\vec{f}_\textmds{F}(\phi)$. 
Here, $y_\textmds{F1}=-y_\textmds{F2}$ and the swimming stroke is parametrised by the beat cycle angle $\phi \in [0,2\pi]$ with a stroke frequency of $53\mbox{Hz}$.
We assume that CR that does not rotate about the $x$ axis, so that $\vec{f}_\textmds{F}$ is purely along the $x$ direction. This assumption can be relaxed straightforwardly. 

The swimmer moves with a velocity $\vec{v}_\textmds{S}(\phi)$ along the $x$ axis, where the velocity is taken from the measurements in Ref. \citep[][Fig. 4b]{Guasto2010}, but reduced by a factor of 0.7 to account for the confinement in our experiments. 
Note that the swimmer velocity oscillates throughout the stroke period, where the speed averaged over a swimming stroke is $\langle \VS \rangle \approx 84 \mummys$.
The distance progressed per stroke is $d_\textmds{S} \approx 2.25 \mummy$ with a forward : backward motion ratio $\approx 3 : 0.8$.

The flow field of the simulated microswimmer is the superposition of the flow from the spherical cell body, $\vec{u}_\textmds{B}$, and that from the two flagella, $\vec{u}_\textmds{F1,2}$. The former is obtained from the well known solution to a no-slip sphere dragged at speed $\vec{v}_\textmds{S}$ through a viscous fluid, which in the co-moving frame is:
\begin{align}
\label{eq:BodyFlowField}
\vec{u}_\textmds{B} (\vec{r}, \phi) 
&= 
\frac{\vec{v}_\textmds{S}}{r} \left( \frac{3 \RS}{4} + \frac{\RS^3}{4 r^2} \right) \nonumber \\ &+
\frac{( \vec{v}_\textmds{S} \cdot \vec{r}) \vec{r} }{r^2} \left( \frac{3 \RS}{4 r} - \frac{3 \RS^3}{4 r^3} \right),
\end{align}
where $\vec{r} = \vec{x} - \vec{x}_\textmds{S}$ 
and $r = |\vec{r}|$.
Unless otherwise mentioned, we will use $\RS = 3.5 \mummy$ for CR throughout this work.

The flagella are modelled as point forces, which in presence of the no-slip spherical body generate the flow  $\vec{u}_\textmds{F1,2} = \underline{\underline{G}}^* \cdot \vec{f}_\textmds{F1,2}/8\pi\eta$. Here $\underline{\underline{G}}^*(\vec{x}_\textmds{F1,2})$ is the Green's function for the no-slip sphere system (see \cite{spagnolie2015geometric}).
The flagellar force $\vec{f}_\textmds{F}$ is then related to the swimmer velocity $\vec{v}_\textmds{S}$ through the requirement of zero net force on the system. This can be imposed by subtracting from the total flagellar force the net force exerted by it on the body   
\begin{align}
\label{eq:ForceIntegral}
\vec{f}_\textmds{F\actson B} &= \oint_S \underline{\underline{\sigma}}_\textmds{F} \cdot \vec{dS},
\\
\underline{\underline{\sigma}}_\textmds{F} &= -p_\textmds{F} \underline{\underline{I}} + \eta (\nabla \vec{u}_\textmds{F} + (\nabla \vec{u}_\textmds{F})^T),
\end{align}
where the integral runs over the surface of the microorganism, and $\underline{\underline{\sigma}}_\textmds{F}$ is the stress tensor of the flagellar flow field. As a result, the net force exerted by the flagellum on the fluid becomes  $\vec{f}_\textmds{F\actson L} = \beta \vec{f}_\textmds{F}$, with 
\begin{align}
\label{eq:ForceBalanceFunction}
\beta &= 1 + \frac{\RS^3 (2\varrho_x^2 - \varrho_y^2)-3\RS \varrho^2(2\varrho_x^2 + \varrho_y^2)}{4 \varrho^{5}},
\end{align}
depending on the relative position between the flagella and the body, $\vec{x}_\textmds{F} - \vec{x}_\textmds{S} = (\varrho_x,\varrho_y,0)$, and the distance $|\vec{x}_\textmds{F} - \vec{x}_\textmds{S}| = \varrho$. Balancing the flagellar and body forces on the fluid one obtains the relation between the instantaneous swimming velocity and the instantaneous flagellar force:
\begin{align}
\label{eq:FlagellarForceSolution}
\frac{\vec{f}_\textmds{F}}{8\pi \eta} &= -\frac{1}{\beta} \frac{3 \RS \vec{v}_\textmds{S}}{8}.
\end{align}

\begin{figure*}[t]
    	\includegraphics[width= \linewidth]{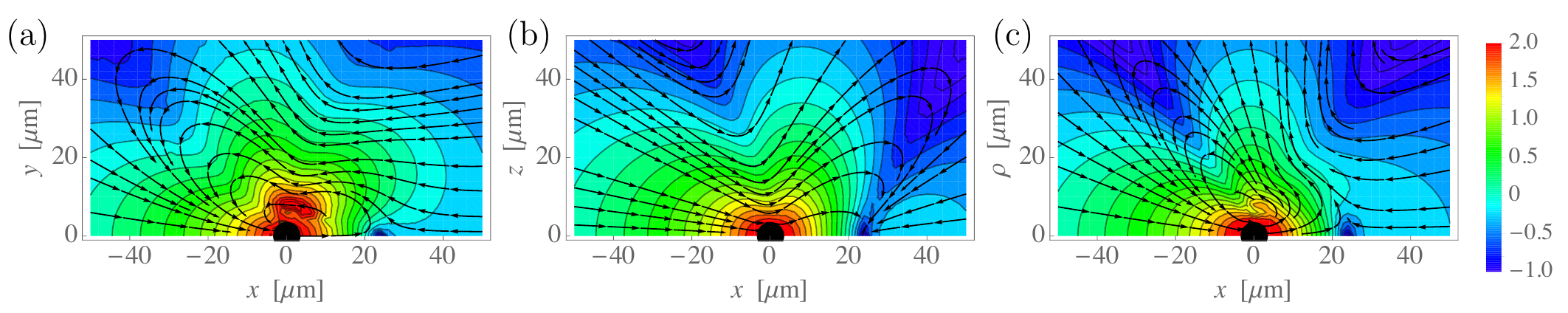} \\
  	\caption{
Flow fields generated by the model \chlamy in the lab frame, time-averaged over one beat cycle.
(a) is the $z=0$ cross section in which the flagella move, 
(b) is the $y=0$ cross section perpendicular to the flagellar plane, and
(c) shows the time- and azimuthally-averaged flow.
The swimmer body (black disc) is oriented in the $x$ direction.
The velocity directions in each plane are shown by streamlines (black arrows) and the magnitude by colours, ranging from $10^{-1} \mummys$ (blue) to $10^{2} \mummys$ (red) on a logarithmic scale.
}
\label{fig:AverageFlowFields}
\end{figure*}

Note that during the simulations of tracer particles near the model swimmer, the singularities of the external Stokeslets are regularised by capping the tracer's advection speed to the organism's speed. 
We have tested simulations with different cut-off values and cut-off descriptions, but we observe these lead to very similar results, because the amount of time a tracer spends close to the Stokeslets is short. 
We do not intend to capture flagellar interactions with great accuracy in these simulations, but we aim to have a good description of the no-slip layer flows close to the body, ensuring the overall flow is force-free.
The swimmer-generated flow field, $\vec{u}_\textmds{S} = \vec{u}_\textmds{B} + \vec{u}_\textmds{F}$, is obtained by combining Eqs. \ref{eq:BodyFlowField}--\ref{eq:FlagellarForceSolution} for a prescribed flagellar trajectory $\vec{x}_\textmds{F1}$.
For CR we use a loop-like trajectory next to and mostly in front of the cell (\fig{fig:Diagram}), which is far from the body during the power stroke ($0 \le \phi \le \pi$) and close to the body during the recovery stroke ($\pi \le \phi \le 2\pi$).
The resulting flow field is shown in \fig{fig:FlowFields} (time-resolved) and  \fig{fig:AverageFlowFields} (time-averaged); and in Supplementary Videos 7 and 8 in the lab frame and rest frame, respectively. These flows compare well with experimental measurements (see \citep[][Fig. 3]{Guasto2010} and  \citep[][Fig. 3a]{Drescher2010f}). They display a characteristic fluctuation, within a beat, between contractile (power stroke) and extensile (recovery stroke) behaviour, with a stagnation point at position  $(x_0,0,0)$ which, in the limit $\RS \to0$, satisfies
\begin{align}
\label{eq:StagnationPointPosition}
\frac{\varrho_y^2 + 2 (x_0-\varrho_x )^2}{\left(\varrho_y ^2+(x_0-\varrho_x )^2\right)^{3/2}}-\frac{2}{x_0} = 0.
\end{align}
By solving this expression numerically, we superimpose the solution for $x_0$ as a red point in \fig{fig:FlowFields} and Video 1 throughout the beat cycle.

\begin{figure*}[t]
    	\includegraphics[width= \linewidth]{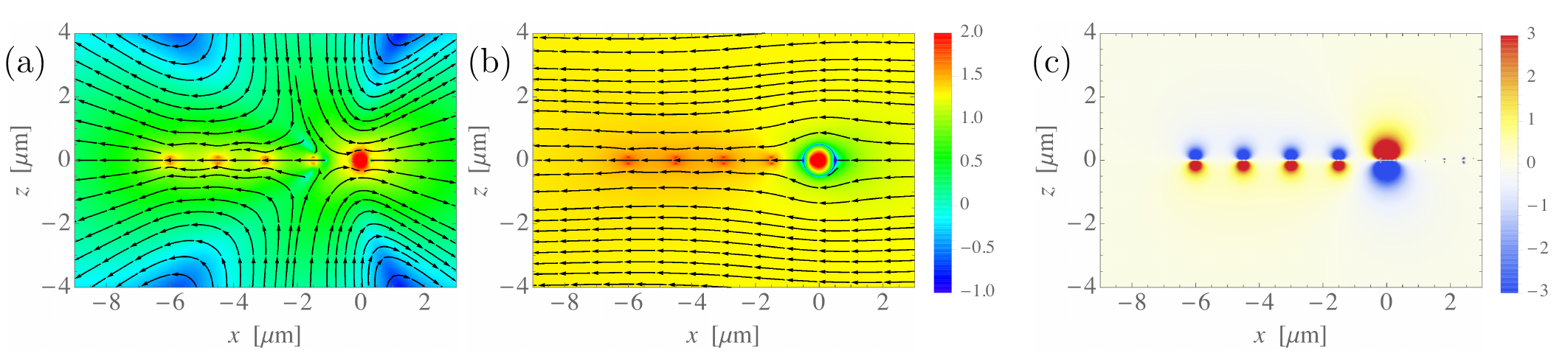} \\
  	\caption{
Flow fields generated by the model \textit{E. coli} bacterium, time-averaged over one helix rotation.
The swimmer body is oriented in the $x$ direction, moving with $\VS = 25 \mummys$ and rotating with $\Omega_\textmds{S}=10 /\mbox{s}$.
(a) Stream lines of the radial and tangential flows in the $(x,y)$ plane at $z=0$, in the laboratory frame. 
Colours portray the flow's magnitude on a logarithmic scale, ranging from $10^{-1} \mummys$ (blue) to $10^{2} \mummys$ (red).
(b) Same, in the co-moving frame. \ar{Notice the flow vanishes at the cell surface, satisfying the no-slip boundary condition}.
(c) Azimuthal flows due to the head-tail counter rotation.
Colours give the magnitude of the flow's $z$ component, on a linear scale, into the board (blue) and out of the board (red).}
\label{fig:EColiFlowFields}
\end{figure*}

\subsection{Model for \textit{E. coli}}
\label{sec:ECflow}

To model the time-averaged flow of the bacterium  \textit{E. coli} (EC), which is  propelled by a rotating and left-handed helical flagellum of length $\sim 6 \mummy$, we consider a body radius $\RS = 0.5 \mummy$, and represent the flagellum with $N$ steady Stokeslets located at $\vec{x}_{\textmds{F}i} = (-i \lambda+x_\textmds{S},0,0)$, where $i \in [1, \dots N]$, $\lambda >\RS$. The body moves with a constant swimming speed $\VS = 25 \mummys$ along the $x$ axis.
The body rotates with a constant angular velocity $\Omega_\textmds{S} = 10 \textmd{Hz}$, balanced by the flagellar counter rotation represented here by four rotlets at the same positions $\vec{x}_{\textmds{F}i}$. The Stokeslet and rotlet intensities,  $\vec{f}_\textmds{F}$ and $\vec{\tau}_\textmds{F}$ are assumed to be the same across the $N$ point forces and torques, and are chosen to guarantee a system with zero net force and torque. The Stokeslet is selected following the reasoning of the previous section, leading to
\begin{align}
\frac{\vec{f}_\textmds{F}}{8\pi \eta} &= - \left( \sum_{i=1}^N \beta^{(i)} \right)^{-1} \frac{3 \RS \vec{v}_\textmds{S}}{4},
\end{align}
where 
\begin{align}
\beta^{(i)} &= 1+ \frac{3 \RS (i \lambda)^2 -\RS^3}{2 (i \lambda)^3}.
\end{align}
To fix the rotlet $\vec{\tau}_\textmds{F}$, start from the rotational flow due to the body \cite{kimmicrohydrodynamics}:
\begin{align}
\label{eq:RotationalBodyFlowField}
\vec{u}_\textmds{R} (\vec{r}) 
&= 
\frac{\RS^3}{r^3} \vec{\Omega}_\textmds{S} \times \vec{r},
\end{align}
where $\vec{\Omega}_\textmds{S} = (\Omega_\textmds{S},0,0)$. The full flagellar flow is given by 
\begin{align}
\label{eq:EColiFlagFlow2}
\vec{u}_{\textmds{F}i} 
&= 
\underline{\underline{G}}^* \cdot \frac{\vec{f}_\textmds{F}}{8\pi\eta} + \underline{\underline{T}}^* \cdot \frac{\vec{\tau}_\textmds{F}}{8\pi\eta},
\quad
\underline{\underline{T}}^* = \frac{1}{2}\bnabla \times \underline{\underline{G}}^*.
\end{align}
The torque exerted by the $i$-th rotlet on the fluid is  $\vec{\tau}^{(i)}_\textmds{F,L} = \gamma^{(i)} \vec{\tau}_\textmds{F}$, where 
\begin{align}
\gamma^{(i)} &= 1+ \frac{\RS^3}{(i \lambda)^3}.
\end{align}
Balancing the total torque on the fluid from the body and the rotlets, $\vec{\tau}_\textmds{F}$ is found to be 
\begin{align}
\frac{\vec{\tau}_\textmds{F}}{8\pi \eta} &= - \left( \sum_{i=1}^N \gamma^{(i)} \right)^{-1} \RS^3 \vec{\Omega}_\textmds{S}.
\end{align}
This fixes the representation \eq{eq:EColiFlagFlow2} of the flow due to the $i$-th component of the discrete flagellum.
Together with the translation \eq{eq:BodyFlowField} and rotation \eq{eq:RotationalBodyFlowField} of the body, this completes the model for EC.
\fig{fig:EColiFlowFields} shows the flow for the model of an \textit{E. coli} bacterium, with $N=4$ flagellar Stokeslets and position $\lambda = 1.5 \mummy$.

\subsection{Model for \textit{Oxyrrhis marina}}
\label{sec:ModelOxy}

Lastly, we focus on the organism \textit{Oxyrrhis marina} (OM), which is a dinoflagellate that propels by beating its posterior flagellum like a sperm cell. 
For simplicity, we use a time-averaged model with body radius $\RS = 9 \mummy$, and the flagellum of length $48 \mummy$ is represented by four steady Stokeslets located at $\vec{x}_\textmds{F1} = (-i \lambda+x_\textmds{S},0,0)$, where $i \in [1,2,3,4]$, $\lambda = 12 \mummy$, and the body moves with constant swimming speed $\VS = 100 \mummys$ along the $x$ axis.

\subsection{Simulating the outboard model}
\label{sec:Simulating}

Once the flow fields $\vec{u}_\textmds{S}(\vec{x},t)$ are known, we can simulate the dynamics of tracer particles under the assumption of straight swimming by the microorganism. This approach is in the spirit of recent work by Shum and Yeomans \cite{shum2017entrainment}.

For each tracer size, $\RT = 10^{-2+3 (i-1)/(8 - 1)} \mummy$ for CR and OM and $\RT = 10^{-2+2 (i-1)/(8 - 1)} \mummy$ for EC, where $i = 1,2,...,7$, simulations are performed with an ensemble of $N=10^3$ particles that do not interact with each other.
Tracers are advected by the swimmer-generated velocity field with a velocity given by the Fax\'en relation,
\begin{align}
 \vec{v}(\vec{x},t) = \left (1+\frac{1}{6} \RT^2 \nabla^2 \right ) \vec{u}_\textmds{S}(\vec{x},t).
\end{align}
Particles experience steric interactions with the swimmer through a hard-core repulsion. 
We tried various prescriptions of the repulsion potential, including the Weeks-Chandler-Anderson potential ($\sim r^{-12}$ for small $r$, with a cut-off radius $r_c\ll\RT$) as well as a softer potential ($\sim r^{-6}$).
To save computation time, it is also possible to set $V=0$ and if the particle overlaps with the swimmer after a timestep, $r<\RS+\RT$ in the co-moving frame, renormalise the distance to $r=\RS+\RT$ but keep the new polar angle $\theta$.
Supplementary Videos 9,10 (laboratory and co-moving frames respectively) \cite{Supplementary}, show the tracers advected by the outboard CR in the deterministic regime.

Brownian motion is simulated with a standard Gaussian white noise $\vec{\xi}$, with a diffusion constant given by the Stokes-Einstein relation $D_0(\RT) = k_B{\rm T}/6 \pi \eta \RT$, where $k_B$ is the Boltzmann constant, ${\rm T}$ is the temperature (not to be confused with the contact time $T$), and $\eta$ is the fluid viscosity.

\section{Contact time theory}
\label{sec:TheoryStreamlines}

\begin{figure}
    	\includegraphics[width= 0.9\linewidth]{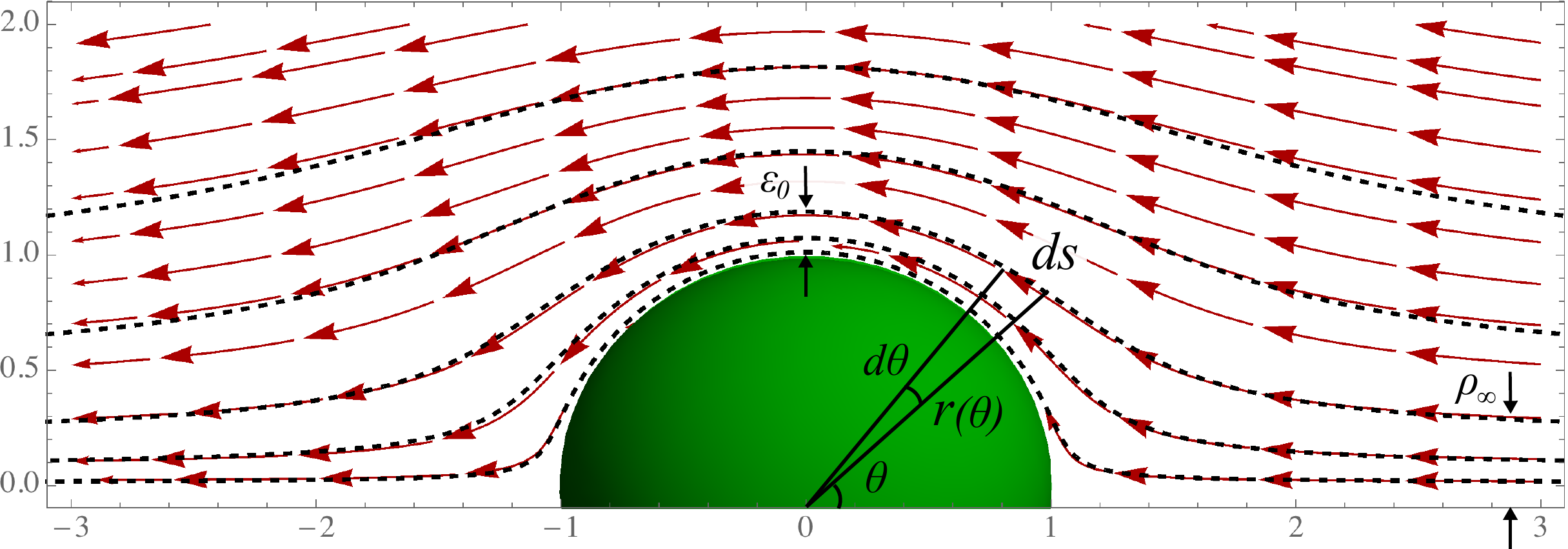} \\
	\includegraphics[width=0.9 \linewidth]{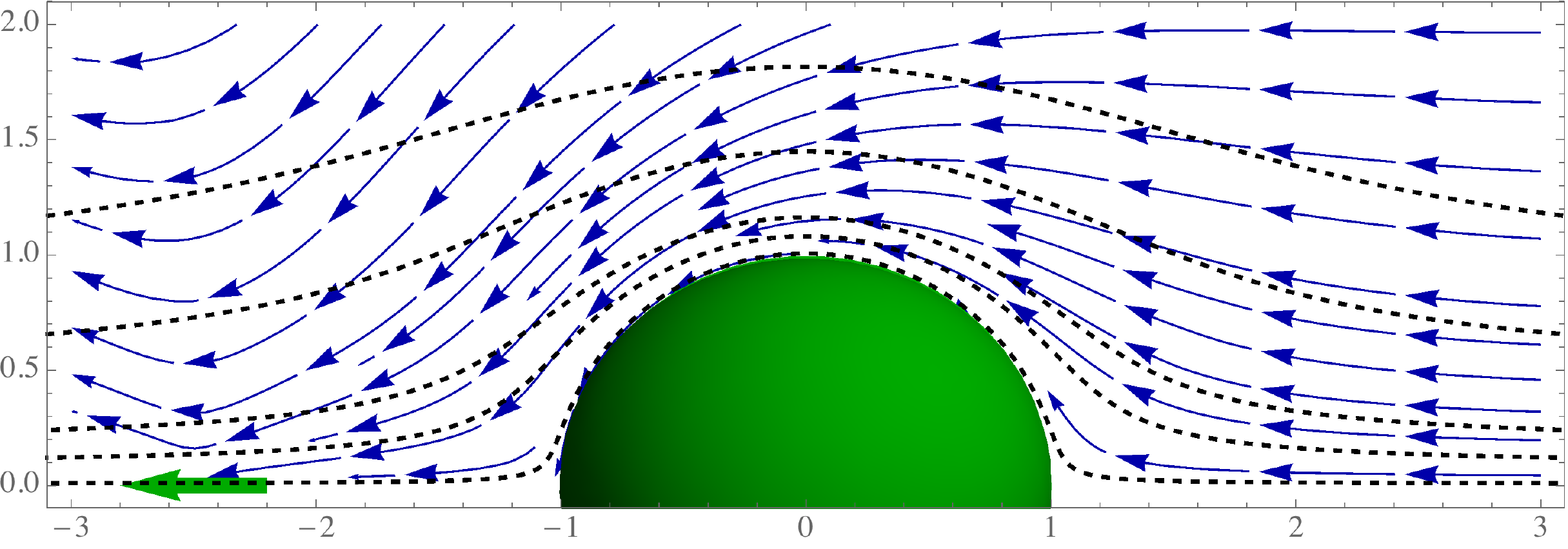}
\caption
[Flow past a micro-swimmer]
{
Streamlines of flow fields generated by a micro-swimmer, in its rest frame.
The organism is oriented along the positive $x$ axis, moves with speed $\VS = 25 \mummys$, has a body of radius $\RS = 1 \mummy$ located at the origin [green sphere], and its flagella are represented by a Stokeslet located at $\lambda = 2.5\mummy$ and pointing in the negative $x$ direction.
Panel (a) shows the flow contribution $\vec{u}^\textmd{B}$ from the swimmer body and panel (b) shows the flow $\vec{u}^\textmd{F}$ due to the flagella [green arrow].
The dashed black lines are approximate stream lines from \eq{eq:BodyStreamLine}.
}
\label{fig:FlowPastSwimmer}
\end{figure}

This section develops a simple estimate for the contact time between a tracer particle of radius $\RT$ and a simplified  `outboard' swimmer. This section will then be complemented by a description of the full distribution of entrainment times in Appendix \ref{sec:DistributionTheory}.
The swimmer is here simplified to a spherical body of radius $\RS$ and a single flagellum represented by one external Stokeslet 
oriented along the negative $x$ direction and located at a position $x = \pm\lambda \RS$ with respect to the body centre ($\lambda > 0$; $+$ for a puller; $-$ for a pusher). The force- and torque-free swimmer moves with a constant velocity $\VS$ along the positive $x$ axis, and we examine the motion of a tracer in the frame co-moving with the swimmer.

\subsection{Flow close to the swimmer's body}

\begin{figure}
    	\includegraphics[width= \linewidth]{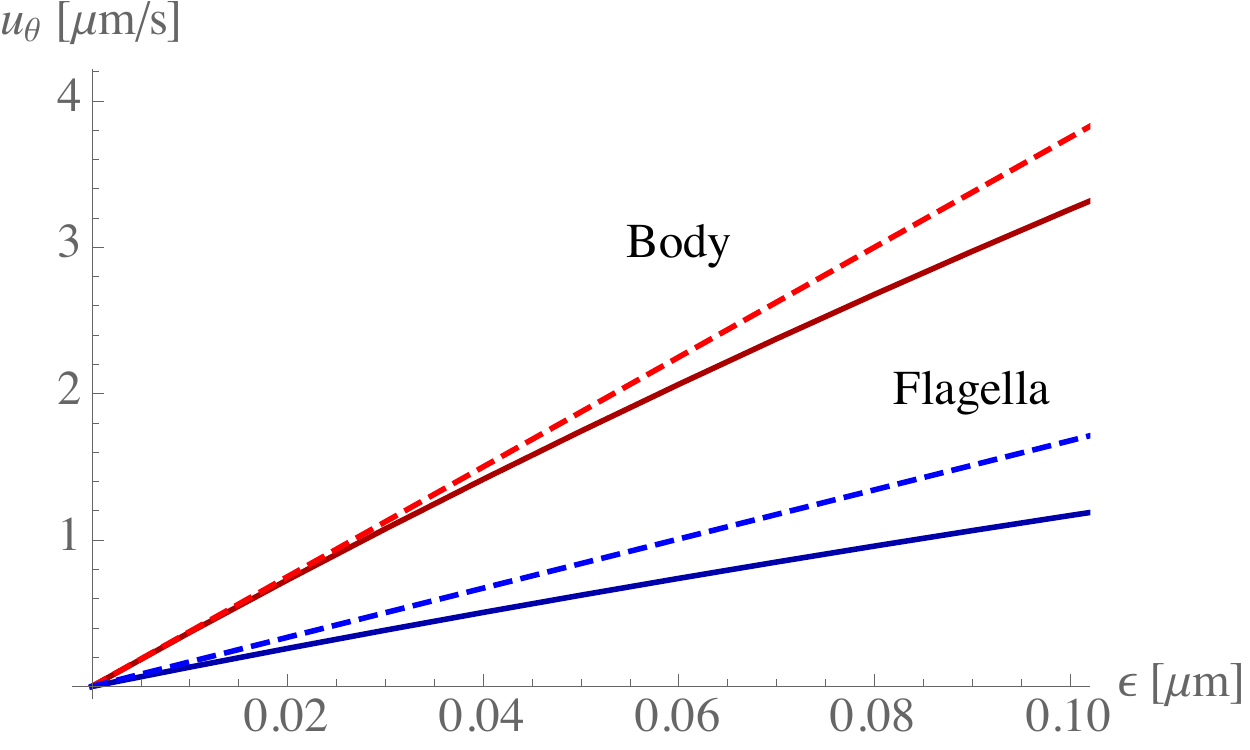}
\caption
[Tangential flow near a micro-swimmer]
{
Magnitude of the tangential flow along the body of a micro-swimmer, defined the same as in the caption of \fig{fig:FlowPastSwimmer}, at the equator $\theta = \pi/2$ as a function of the distance from the body $\epsilon$.
The components due to the body and flagella are shown with red lines and blue lines, respectively. 
Dashed lines are the linear approximations given by Eqs. \ref{eq:BodyTangential}--\ref{eq:FlagellaTangential}.
}
\label{fig:FlowPastSwimmerNear}
\end{figure}

The swimmer-generated flow field is the sum of the body and flagellar flows: $\vec{u}_\textmds{B}$, and $\vec{u}_\textmds{F}$ (see Appendix \ref{sec:ModelChlamy}).
The body flow, \eq{eq:BodyFlowField}, can be obtained from the stream function
\begin{align}
\label{eq:BodyStreamFunction}
\psi_\textmds{B}(r, \theta) &= - \frac{\VS}{2} \left( r^2 + \frac{\RS^3}{2r} - \frac{3 \RS r}{2} \right) \sin^2 \theta.
\end{align}
This provides the streamlines for a given impact parameter $b$:
\begin{align}
\label{eq:BodyStreamLine}
r_\textmds{SL}(\theta) &= \RS + \sqrt{\frac{2}{3}} \frac{\impar}{\sin \theta} + \mathcal{O}\left( \frac{\impar}{\RS} \right)^2,
\end{align}
where the closest distance of approach is given by $\epsilon_{0} = \impar \sqrt{2/3} =r_{\textmds {SL}}(\pi/2)-\RS$ at the polar angle $\theta = \pi/2$;  as well as the flow field 
\begin{align}
\label{eq:BodyTangential}
u_\theta^\textmds{B}(r,\theta) &= \frac{3 \epsilon \VS}{2 \RS} \sin \theta + \mathcal{O}\left( \epsilon^2 \right),
\end{align}
where $\epsilon = r - \RS$. Figures~\ref{fig:FlowPastSwimmer},\ref{fig:FlowPastSwimmerNear} show that both expressions are valid close to the swimmer's body. Note that this is a pusher-type swimmer, but the same derivation holds for pullers.
Moreover, Fig.~\ref{fig:FlowPastSwimmer}b) shows that the functional shape in \eq{eq:BodyStreamLine} is also a good estimate for the flagellar streamlines close to the swimmer body. The flagellar flow can be expanded in this region as
\begin{align}
\label{eq:FlagellaTangential}
u_\theta^\textmds{F} &= \frac{9 \lambda^3(1+\lambda)^2}{2 (1+2\lambda)(1+\lambda^2)^{5/2}} \frac{\epsilon \VS}{\RS} \sin \theta + \mathcal{O}\left( \epsilon^2 \right).
\end{align}
The flow field approximations Eq.~\ref{eq:BodyTangential},\ref{eq:FlagellaTangential} are shown in \fig{fig:FlowPastSwimmerNear} with dashed lines, compared to the exact flows with solid lines, at the swimmer's equator $\theta = \pi/2$. 
Note that both approximate flows satisfy the no-slip boundary condition, and grow linearly with $\epsilon$ close to the body. 
For $\lambda \sim \RS$, $u_\theta^\textmds{F}$ is of the same order of magnitude than $u_\theta^\textmds{B}$, whereas in the limit $\lambda \to \infty$, $u_\theta^\textmds{F}\sim1/\lambda$ and we recover the body flow as that generated by a sphere dragged through the liquid.
Overall, the total tangential flow near the body in the co-moving frame can be written as
\begin{align}
u_\theta(r, \theta)&=u_\theta^\textmds{B}+u_\theta^\textmds{F}
\\
\label{eq:Tangential}
&= \frac{3 \VS \epsilon}{2 \RS g(\lambda)} \sin \theta + \mathcal{O}\left( \epsilon^2 \right),
\\
g(\lambda) &= \left(1 + \frac{3 \lambda^3(1+\lambda)^2}{(1+2\lambda)(1+\lambda^2)^{5/2}} \right)^{-1}.
\end{align}
For a more general flagellar orientation and position we still expect a similar functional form but with a more complex function $g(\vec{r}_\textmd{f}) = (1+ \tilde{g})^{-1}<1$, because the flagella always act to push or pull the liquid past the swimmer body faster, averaged over a swimming stroke.

\begin{figure*}
\centering
	\includegraphics[width=1\linewidth]{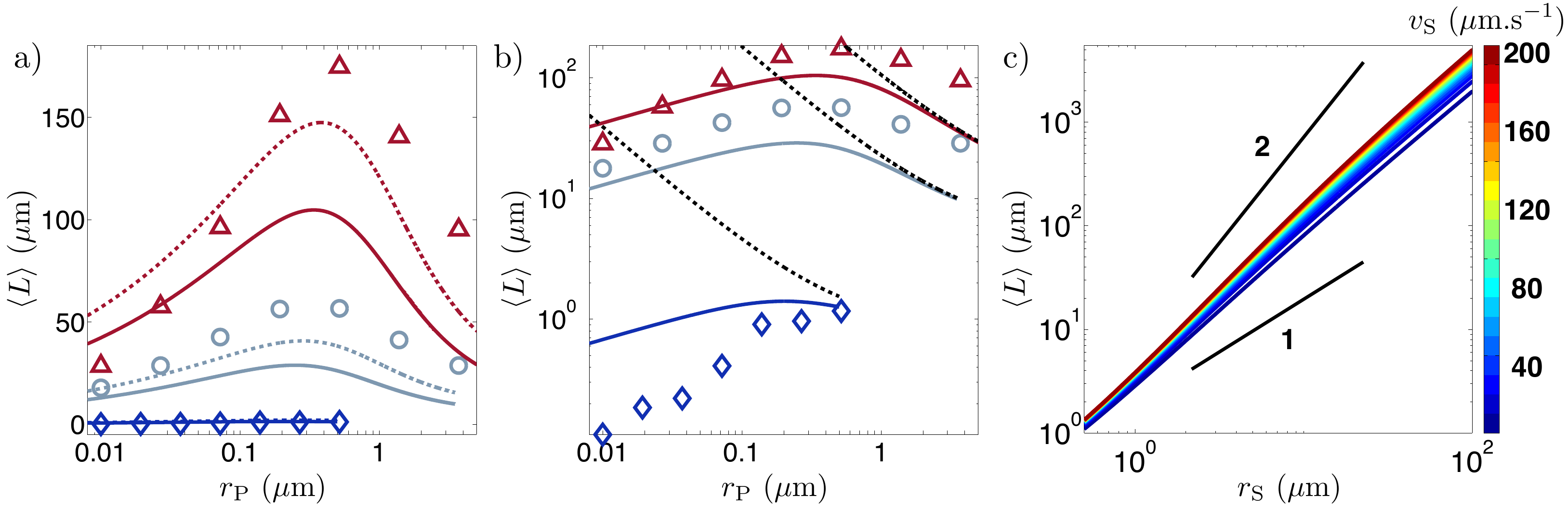}
  	\caption{a) Comparison of the entrainment lengths obtained by taking in \eq{eq:StochasticEOM1} $U=\langle U(x)\rangle_{x}=3 \VS /(\pi\RS g(\lambda))$ (dashed lines) or $U=3 \VS /(2\RS g(\lambda))$ (full lines). The first approach gives better quantitative agreements with the numerical simulations. b) Same plot as in Fig. 3a-main text with a log-log scale. The entrainment by a bacterium (blue diamonds) never exceeds $\sim 1 {\rm \mu m}$ because of its small size. c) Evolution of the entrainment length with the size of the organism for a fixed particle size ($r_{\rm P}=0.5 {\rm \mu m}$) at different velocities $v_{\rm S}$ (colorbar) and fixed $\lambda=4$. The bigger the organism, the longer the interaction.}
\label{figureSI2}
\end{figure*}

\subsection{Contact time without noise}

Our experiments show that a particle is entrained further by an organism if it spends more time near its cell wall, in the `no-slip layer'. 
We compute the time taken for a particle to be advected along a stream line (\eq{eq:BodyStreamLine}) close to the swimmer
\begin{align}
\label{eq:EntrainmentTime1}
T &\approx \int_\textmds{SL} \frac{ds}{u_\theta} = \int_0^\pi \frac{ds}{d \theta} \frac{1}{u_\theta [r_\textmds{SL}(\theta),\theta]}d\theta,
\end{align}
where $ds$ is the arclength differential along the stream line parameterised by the angle $\theta \in [0,\pi]$.
Inserting \eq{eq:BodyStreamLine} into \ref{eq:Tangential}, we find that the tangential flow along a streamline is constant to first order:
For a given impact parameter $\impar$ we have 
\begin{align}
\label{eq:Tangential2}
u_\theta [r_\textmds{SL}(\impar)] &=\frac{3\VS \epsilon_{0} }{2 \RS}\frac{1}{ g(\lambda)} + \mathcal{O}\left( \epsilon_0^2 \right),
\end{align}
where $\epsilon_{0}=\impar \sqrt{2/3}$ is the closest distance of approach.
Therefore, taking the stream line length as the distance that a particle must travel around the swimmer, $S = \int_\textmds{SL} ds = \pi (\RS+\RT)$ for small impact parameters, yields the contact time
\begin{align}
\label{eq:EntrainmentTime2}
T &= \frac{2 \pi \RS (\RS+\RT) }{3\VS \epsilon_{0} } ~g(\lambda)
\\
&= \frac{ \sqrt{2/3} \pi \RS (\RS+\RT) }{\VS \impar} ~g(\lambda).
\end{align}
Note that in the limit of point-like tracer particles, $\RT \to 0$, we recover the result by Mueller and Thiffeault \cite{mueller2017fluid}: the entrainment length $L = \VS T = C \RS^2/\impar$, where the constant $C = \sqrt{2/3} \pi \sim 2.565$ for a bare no-slip sphere, as $\lambda \to \infty$ such that $g(\lambda) \to 1$, and $C < 2$ for typical swimmers.
A particular feature of interest here are that the contact time, and hence the entrainment length, increases quadratically with the swimmer size $\RS$. 
This implies there is a large difference between \textit{E. coli} bacteria and \textit{Chlamydomonas} algae.

Finite-sized particles do not have access to the stream lines very close to the swimmer body. 
Therefore, if the impact parameter is so small that $0<\epsilon_{0} < \RT$, these particles collide with the front of the swimmer body.
Lubrication forces and steric interactions expel them, so that they cross stream lines. 
Hence, in the co-moving frame, they approximately move along a circular trajectory around the body when $0<\theta < \pi/2$ and into the orbit of closest approach, the streamline with $\epsilon_{0} = \RT$, at $\theta = \pi/2$.
As a result, the average contact time is reduced for large particles, because they cannot reach the no-slip layer close to the body and thus flow past the swimmer, as seen in its co-moving frame, more quickly. 

Finally, we see that the solution diverges as $1/\impar$ with decreasing impact parameter.
This can lead to very long entrainment events, like a ball pushed on the nose of a seal.
However, this position is unstable with the introduction of thermal fluctuations, as we consider next.

\subsection{Contact time with noise}
\label{subsec:ContactTimeNoise}

We consider a Brownian particle advected in a linear shear flow over a straight solid surface that mimicks the swimmer's cell wall (main text Fig. 4a).
The flow velocity is $\vec{u} = \epsilon U  \vec{e}_x$, where the strain rate $U$ derives from the velocity along a streamline, using \eq{eq:Tangential2}, so that 
\begin{align}
\label{eq:RateOfStrain1}
U=3 \VS/(2\RS g(\lambda)).
\end{align}
A particle of radius $\RT$ is initially positioned at $(x=0,\epsilon=\RT)$, disperses with diffusivity $D$ and is advected by the flow $\vec{u}(\epsilon)$, but cannot cross the line $\epsilon=\RT$. 
Without loss of generality \cite{vankampen1983} this system is mapped to an `image' system (main text Fig. ¡4b). 
Here, the particle is initially located at $(x=0,y=0)$, the modified flow is $\vec{v} = (\RT + |y|)U \vec{e}_x$, and the tracer \textit{can} cross the surface.
We aim to compute the average time $\langle T \rangle$ needed for the colloid to travel a distance $S = \pi(\RS+\RT)$ along the positive $x$-direction.
The stochastic equations of motion are
\begin{align}
\label{eq:StochasticEOM1}
\dot{x}(t) = \big(\RT + |y| \big)U + \xi_x(t), \quad
\dot{y}(t) = \xi_y(t),
\end{align}
where the noise correlations are defined as
\begin{align}
\label{eq:StochasticEOM2}
\langle \xi_i \rangle = 0, \quad \langle \xi_i (t) \xi_j (t')\rangle = 2D \delta_{ij} \delta(t-t').
\end{align}
Note that the Fax\'en correction for finite-sized tracer particles need not be included here, as the Laplacian acting on pure shear flows vanishes. 
Integrating and averaging \eq{eq:StochasticEOM1} gives
\begin{align}
\label{eq:MeanPositionX1}
\langle x(t) \rangle  
&= \int_0^t dt' \big(\RT + \langle |y(t')| \rangle \big)U + \langle \xi_x(t')\rangle
\\
&= \RT U t + \int_0^t dt' \left \langle \left | \int_0^{t'} dt'' \xi_y(t'') \right | \right \rangle
\\
\label{eq:MeanPositionX1c}
&= \RT U t + \int_0^t dt' \langle |y(t')| \rangle.
\end{align}
Using the initial condition that particles start from $y=0$, we can employ the canonical distribution $p(y,t) = e^{-y^2/4Dt} / \sqrt{4 \pi D t}$ to give
\begin{align}
\label{eq:MeanPositionY}
\langle |y(t')| \rangle = \int_{-\infty}^\infty |y| ~p(y,t') dy = \sqrt{\frac{4Dt'}{\pi}}.
\end{align}
Inserting this expression into \eq{eq:MeanPositionX1c} and integrating once more then yields
\begin{align}
\label{eq:MeanPositionX2}
\langle x(t) \rangle  
&= \RT U t + \frac{4 \sqrt{D}}{3 \sqrt{\pi}} U t^{3/2}.
\end{align}
Requiring that $\langle x(T) \rangle = \pi(\RS+\RT)$, we find the mean time $\langle T \rangle$ is as the solution of the cubic equation
\begin{align}
\label{eq:MeanTimeCubic}
0 &= c_0 + c_2 \langle T \rangle + c_3 \langle T \rangle^{3/2},
\\
\label{eq:MeanTimeCubic2}
c_0 &= - \frac{2\pi \RS (\RS+\RT) g(\lambda)}{3 \VS \RT} ;~
c_2 = 1;~
c_3 = \frac{4 \sqrt{D}}{3 \RT \sqrt{\pi}},
\end{align}
which is solved using the Cardano formula. Only one positive and real root exists for all physical situations, $c_0 < 0$ and $c_3 >0$, which is the average contact time
\begin{align}
\label{eq:Cardano}
\langle T \rangle &= \left( C_+ + C_{-} - \frac{c_2}{3 c_3} \right)^2,
\\
\label{eq:Cardano2}
C_\pm &= \sqrt[3]{r_\textmd{C} \pm \sqrt{q_\textmd{C}^3 + r_\textmd{C}^2}}, 
\\
\label{eq:Cardano3}
q_\textmd{C} &= - \frac{c_2^2}{9c_3^2}, ~
r_\textmd{C} = \frac{- 27 c_0 c_3^2 - 2 c_2^3}{54 c_3^3}.
\end{align}
Hence, we can estimate the average entrainment length $\langle L \rangle = \VS \langle T \rangle$. 
This analytical expression can be evaluated for different tracer sizes,  swimmer speed or size, temperature, fluid viscosity, etc. 
Note that the theory is expected to hold best near the optimal tracer size because the approximation $\epsilon \ll \RS$ in equations (\ref{eq:BodyStreamLine}--\ref{eq:Cardano3}) holds best for particles that follow paths close to the body.

\subsection{Streamline crossing}

\begin{figure}
\centering
	\includegraphics[width=\linewidth]{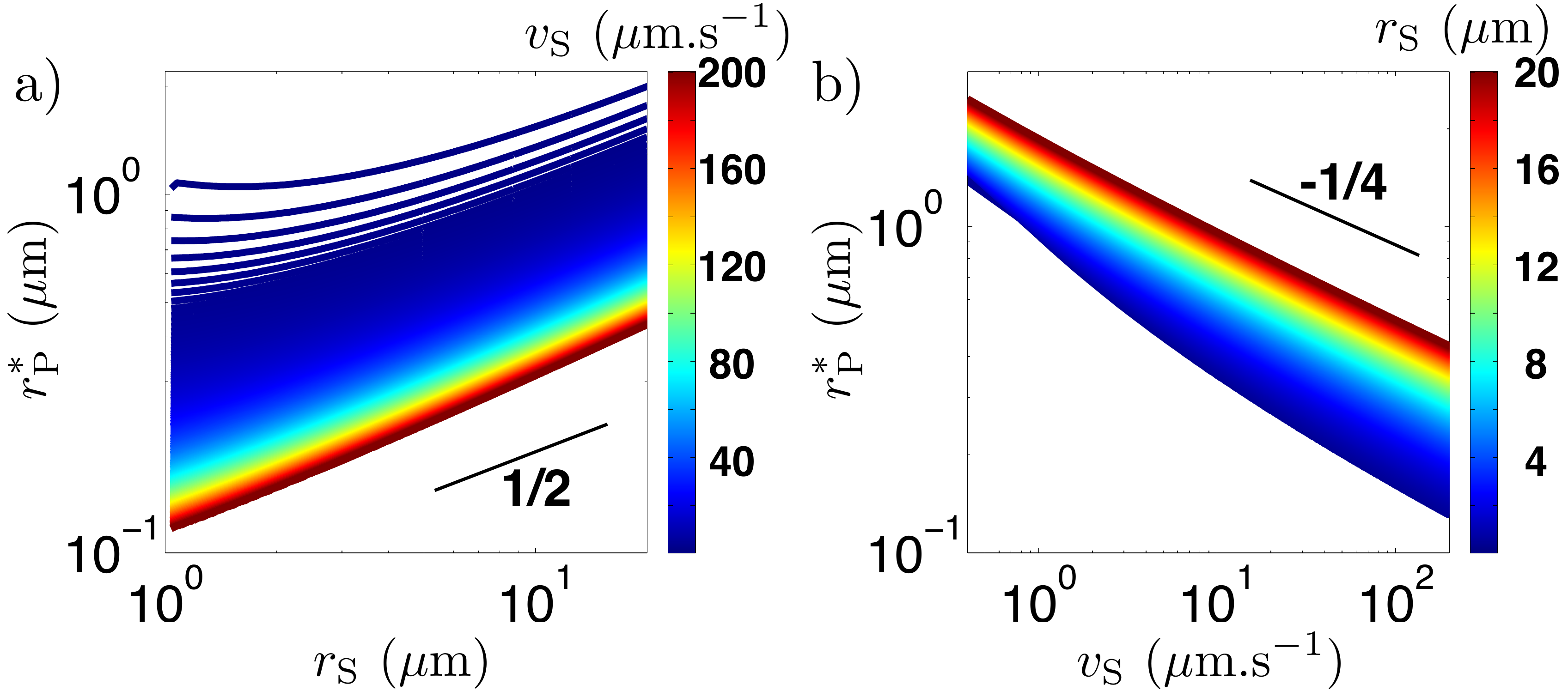}
  	\caption{a) (resp. b)) Evolution of the optimum tracer size $r_{\rm P}^*$ for entrainment with the swimmer's radius (resp. swimmer's speed) at different constant velocities (resp. radii) (colorbar) and constant $\lambda=4$. The increase (resp. decrease) is consistent with a power law dependency with exponent $1/2$ (resp. $-1/4$) as given by our simple estimate from the P\'eclet number (Eq. 9-main text).}
\label{figureSI3}
\end{figure}

In the deterministic limit, $D \to 0$, we recover \eq{eq:EntrainmentTime2} immediately from \eq{eq:MeanTimeCubic}, in agreement with Mueller and Thiffeault \cite{mueller2017fluid}.
In this limit we do not observe a maximum in entrainment length or contact time, but a monotonic decrease with increasing particle size. 
This is because the smaller particles not diffuse away from small $\rho$  values and can access the streamlines closest to the no-slip surface, whereas larger particles cannot access this region due to steric interactions and experience therefore stronger tangential flows on average. 

As a consequence, large particles with small impact parameters, $\impar< \sqrt{3/2}\RT$ i.e. $\epsilon_0<\RT$, can therefore not stay in their original streamline, but must \textit{cross streamlines}.
They move around the swimmer body at distance $r \approx \RT+\RS$ and polar angles $0 \lesssim \theta \lesssim \pi / 2$, and subsequently move along the streamline defined by $\epsilon_0=\RT$  at polar angles $\theta > \pi / 2$ (as seen in main text Fig.~2c-d).
During the first part of this trajectory, according to \eq{eq:Tangential}, these particles experience a tangential flow
\begin{align}
u_\theta(\theta)
&\approx \frac{3 \VS (\RT+\RS)}{2\RS g(\lambda)} \sin (\theta ),
\end{align}
which has an explicit $\theta$ dependence, whereas the flow speed along a streamline (\ref{eq:Tangential2}) during the second part of the trajectory is approximately constant to first order.
Hence, with limited diffusion, large particles might temporarily be `trapped' in the region $\theta \sim 0$ where $u_\theta(\theta) \sim \theta$.
This effect is expected to \textit{increase the contact time}, because next to the time required to flow around the body, one must add the (first-passage) time required to escape the initial low-flow region.

\begin{figure*}[ht]
\centering
\includegraphics[width=0.9\linewidth]{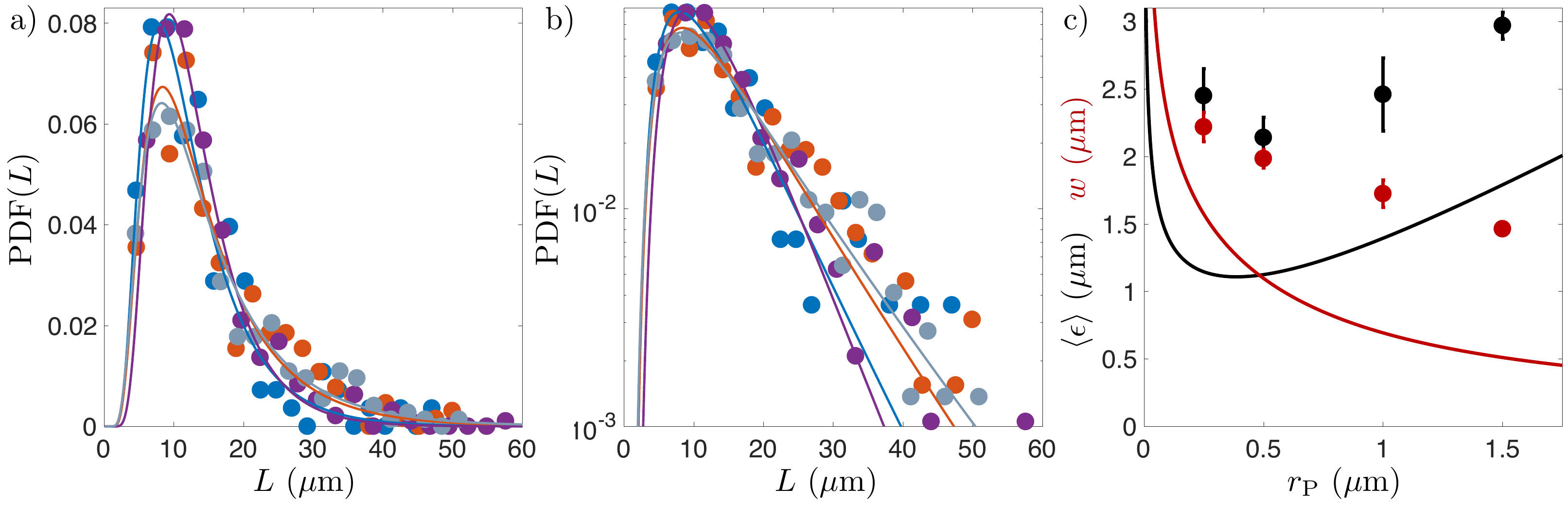}
\caption{a) and b) Plot of the experimental entrainment length distribution obtained with CR for the four tracer sizes probed $\RT=0.25, 0.5, 1$ and $1.5 {\rm \mu m}$ (blue, grey, orange and purple respectively). The full lines represent the fits from Eq. \ref{entr_length} showing an astonishing agreement. c) Comparison between the parameters $\langle \epsilon\rangle$ (black) and $w$ (red) obtained from the fitting procedure (circles) and from the analytical estimation (squares). The agreement is quantitatively reasonable, and the evolution with tracer size is well captured.}
\label{figure10}
\end{figure*}

As a result, our simple theory \textit{underestimates} the contact times observed in our simulations (see main text Fig.~3a). 
Whereas the shape of the curve and position of the maximum are captured well, we see a quantitative difference of a factor of $\sim 2$ near the optimal tracer size.
This is expected, because the impact parameter in the stochastic model ($y(0)=0$)  is effectively equal to $\impar= \RT\sqrt{3/2}$ rather than zero in the simulations.
Indeed, by considering impact parameters in the range $\impar \in [0,\RS]$ instead of $\impar=0$ in our simulations, we retain the shape of the curve $\langle L \rangle (\RT)$ but the magnitude at the maximum is reduced (see main text Fig.~3b).

The current theory hinges on the simplicty of a constant flow speed along a stream line, which allows the stochastic model to be solved with an analytical expression in closed form.
A more accurate approach should account for streamline crossing due to steric interactions, for example with space-dependent strain rate in \eq{eq:StochasticEOM1},
\begin{align}
U(x) &= \frac{3\VS}{2\RS } \frac{1}{g(\lambda)} \sin \left ( \frac{x}{\RS+\RT} \right ),
\end{align}
This non-linear stochastic model is rather complex to solve mathematically and we have not yet explored it in detail. 
However, a way to proceed is to take the spatial average,
\begin{align}
\label{eq:RateOfStrain2}
 U = \langle U(x)\rangle_{x} &= \frac{3\VS}{\pi \RS } \frac{1}{g(\lambda)}.
\end{align}
This gives a constant strain rate, so the model can be solved exactly. The resulting expression is depicted by the dashed lines in Fig. \ref{figureSI2}a. Indeed, this provides a better quantitative agreement with the outboard model simulations, especially for large particles that are subject to streamline crossing.

\section{Entrainment distribution theory}
\label{sec:DistributionTheory}

It is possible to derive an analytical estimate for the distribution of contact times (and therefore entrainment lengths) by approximating the problem as a first-passage time process for a 1D advection-diffusion system.
This is analytically tractable, gives valuable insights, and provides very good quantitative predictions for the distribution of entrainment length (see Fig. 1f main text). 
We start by considering a 1D system where a Brownian particle of diffusivity  $D_{\rm eff}$ is subject to a background drift $V_{\rm eff}$ (along the positive $x$ direction). The particle starts at $x=0$ and we are interested in the probability distribution of the first-passage time $T$ at a boundary $x=S$ where $S>0$. 

The distribution is known to be (see e.g. page 88 in ref. \cite{rednerfirstpassage})
\begin{align}
\label{hitting_times}
{\rm PDF}(T)=\frac{S}{\sqrt{4 \pi D_\text{eff} T^3}}\exp{\left(-\frac{(S-V_\text{eff} T)^2}{4 D_\text{eff} T}\right)}.
\end{align}
If we now consider this particle as the tracer being entrained by the microorganism, the first passage time $T$ can be translated into an entrainment length $L=v_{\rm S}T$ (Eq. 1 main text). This leads to the following distribution of the entrainment lengths:
\begin{align}
\label{entr_length}
{\rm PDF}(L)=\frac{S}{\sqrt{\frac{4\pi D_\text{eff}}{v_{\rm S}} L^3}}\exp{\left(-\frac{v_{\rm S} \left( S-\frac{V_\text{eff}}{v_{\rm S}}L \right)^2}{4 D_\text{eff} L}\right)},
\end{align}
where $S$ is the length that a particle travels around the swimmer, i.e. $S=\pi(r_{\rm S}+r_{\rm P})$. 
Figures~\ref{figure10}a,b and Fig.~1f show that the functional form in Eq.~\ref{entr_length} provides a remarkable fit to all experimental entrainment length distributions, when $V_{\text{eff}}$ and $D_{\text{eff}}$ are kept as fitting parameters. This suggests that the entrainment process can be accurately understood as an effective 1D drift-diffusion process. The challenge is now to relate the effective parameters to the experimental system.

\begin{figure}
\centering
\includegraphics[width=0.7\linewidth]{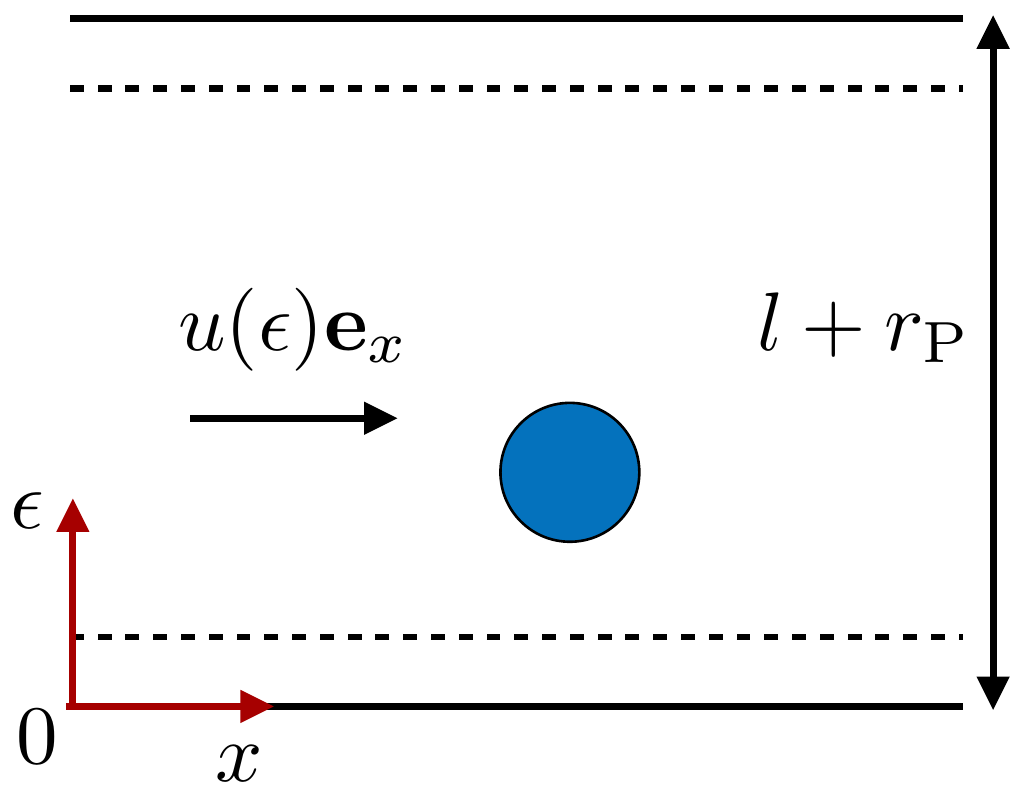}
\caption{Schematic of the geometry considered in the computation of the effective diffusivity $D_{\rm eff}$ through Taylor dispersion.}
\label{figure11}
\end{figure}

To this end, we represent the entrainment process through the 2D system described in \fig{figure11} (see also Fig.~ 4a main text). A Brownian particle of radius $r_{\rm P}$ diffuses with thermal diffusivity $D_0$ in the $x\epsilon$ plane, above an impenetrable wall at $\epsilon=0$; and it is subjected to a uniform background velocity $\mathbf{u}(x,\epsilon,t)=U\epsilon\,\mathbf{e}_x$, with strain rate $U=3v_{\rm S}/(2r_{\rm S}g(\lambda))$. A particle that starts at $(x_0=0,\epsilon_0=\RT)$ will drift with an ensemble-averaged velocity $V_{\rm eff}$ which, after the time $\langle T\rangle$ to cross the length $S=\pi(r_{\rm S}+r_{\rm P})$ (Eq.~\ref{eq:Cardano}), is given by
\begin{align}
\label{effective_velo}
V_{\rm eff}(\langle T\rangle)&=\int u(\epsilon)p(\epsilon,\langle T\rangle)d\epsilon \\
&=U\int \epsilon p(\epsilon,\langle T\rangle)d\epsilon \\
&=U\langle \epsilon\rangle,
\end{align}
where $p(\epsilon,t)$ is the probability of finding the particle at a distance $\epsilon$ from the bottom wall at time $t$. The fitted values of $V_\text{eff}$ can therefore be converted into experimental estimates for $\langle\epsilon\rangle$, and are plotted in Fig.~\ref{figure10}c as black circles. Notice the minimum in the experimental values of $\langle\epsilon\rangle$, which corresponds to the optimal tracer size for entrainment. At the same time, the function $p(\epsilon,t)$ is well known and therefore we can estimate $\langle\epsilon\rangle$ as
\begin{align}
\label{epsilon_ave}
\langle\epsilon\rangle&=\sqrt{\frac{4D_0\langle T\rangle}{\pi}}+r_{\rm P},
\end{align}
which leads to the solid black line in Fig. \ref{figure10}c when using the experimental CR parameters  $\RS=4.5 {\rm \mu m}$, $\VS=49.1 {\rm \mu m.s^{-1}}$, $g(\lambda)=0.69$ (value for the outboard CR), and $D_0=\tilde{D}_0/\RT$.
While quantitatively the expression \eq{epsilon_ave} underestimates the experimental values by $\sim1\,\mu$m, it nevertheless recovers very well the evolution with tracer size $\RT$. In particular, the minima in both cases appear at the same $\RT^*$, and therefore \eq{epsilon_ave} provides a good estimate for the optimal particle size for entrainment. 

The parameter $D_{\rm eff}$ in Eq. \ref{entr_length} can be interpreted as the effective diffusivity of the particles along the velocity direction $x$ (in the 2D process, Fig. 4a and b). We expect this quantity to be enhanced compared to thermal diffusivity $D_0$ because of the background shear flow, as originally realised by Taylor for molecular diffusion within a pipe. Adapting the derivation in \cite{james03} to the case of a uniform shear flow within a thickness $l+\RT$, it is easy to show that the Brownian particles have an effective diffusivity $D_{\rm eff}$ along the $x$-axis given by 
\begin{equation}
D_{\rm eff}=D_0\left(1+ \frac{U^2w^4}{120D_0^2}\right),
\label{eq:effectiveDiff}
\end{equation}
where $w=l-\RT$. The fitted experimental values of $w$ are shown in \fig{figure10}c as red circles.
The length scale $l$ has been introduced in order to reduce the 2D system into an effective 1D process, and  in terms of the entrainment process, it can be interpreted as the transverse length scale the beads explore before reaching the end of the body at $S=\pi(\RS+\RT)$. We therefore estimate it as the sum of the average position over the surface and an excess due to fluctuations, given by the square root of the variance of the $\epsilon$ distribution:
\begin{equation}
l=\langle \epsilon\rangle+\sqrt{\langle \epsilon^2\rangle-\langle \epsilon\rangle^2},
\label{eq:param_l}
\end{equation}
evaluated at the mean contact time $\langle T\rangle$.
This can be estimated analytically, and for initial condition $\epsilon_0=\RT$ one obtains (see Eq.~\ref{eq:MeanPositionY})
\begin{align}
w &= \sqrt{\frac{4D_0\langle T\rangle}{\pi}}\Bigg(1+\sqrt{\frac{\pi}{2}-1}\Bigg).
\label{param_w}
\end{align}
This estimate is shown as a solid red line in Fig.~\ref{figure10}c: it  underestimates the fitted value of $w$, but captures well the qualitative dependence on tracer size. 
The estimates in \eq{epsilon_ave} and \eq{param_w} could be improved by generalising \eq{eq:MeanPositionX2} for an arbitrary initial position, and then averaging over all impact parameters, $V_{\rm eff}=\langle V(\langle T\rangle(y_0); y_0)\rangle_{y_0}$. However, the results are not analytically tractable and therefore loose the simplicity of the minimal-model approach we focus on here, which already provides a remarkable semi-quantitative description of the dynamics.

\bibliography{Biblio_entrainment}

\end{document}